## SPACE SCIENCES

# Modeling the inner part of the jet in M87: Confronting jet morphology with theory

Hai Yang[1,2,3], Feng Yuan[1,4,2]*, Hui Li[5], Yosuke Mizuno[3,6,7], Fan Guo[5], Rusen Lu[1,8,9], Luis C. Ho[10,11], Xi Lin[1,2], Andrzej A. Zdziarski[12], Jieshuang Wang[13]

The formation of jets in black hole accretion systems is a long-standing problem. It has been proposed that a jet can be formed by extracting the rotation energy of the black hole ("BZ-jet") or the accretion flow ("disk-jet"). While both models can produce collimated relativistic outflows, neither has successfully explained the observed jet morphology. By using general relativistic magnetohydrodynamic simulations and considering nonthermal electrons accelerated by magnetic reconnection that is likely driven by magnetic eruption in the underlying accretion flow, we obtain images by radiative transfer calculations and compared them to millimeter observations of the jet in M87. We find that the BZ-jet originating from a magnetically arrested disk around a high-spin black hole can well reproduce the jet morphology, including its width and limb-brightening feature.

## INTRODUCTION

How a jet forms from a black hole accretion flow has been an important unsolved problem. The current consensus is that jet formation requires a combination of magnetic fields and rotation. The most influential model (BZ-jet) is that of Blandford-Znajek (1), in which the jet is formed by extracting the spin energy of the black hole via the magnetic field lines connected to its event horizon. The analytical predictions of this scenario have been confirmed by general relativistic magnetohydrodynamic (GRMHD) simulations that successfully produce relativistic collimated outflows (2–5). However, it has been unclear whether the BZ-jet can explain the observed morphology of the jet, including its elongated structure, width, and limb-brightening.

As an alternative to black hole spin, the jet can also be powered by extracting the rotation energy of the accretion flow by the magnetic centrifugal force or the gradient of the magnetic pressure (6, 7). Such a disk-jet model has also received support from numerical simulations (8–10). Which of these two leading competing scenarios matches the jets observed in active galactic nuclei?

## RESULTS AND DISCUSSION

This work uses three-dimensional GRMHD simulations to reproduce the detailed structure of the jet in M87 to test its formation mechanism. We simulate a hot accretion flow around a black hole using the ATHENA++ code (11). The details of how we perform the simulations and the numerical convergence test are presented in the "Three-dimensional GRMHD numerical simulation of a hot accretion flow" and "Convergence of numerical resolution" sections in Materials and Methods, respectively. Because the black hole in M87 is likely rapidly rotating and the accretion flow is likely to be a magnetically arrested disk (MAD) instead of undergoing standard and normal evolution (SANE) (12, 13), our fiducial model is a MAD around black holes with spin $a = 0.98$ (MAD98). For comparison, we also simulate a SANE around a black hole with $a = 0.98$ (SANE98) and two MADs around black holes with $a = 0.5$ (MAD05) and $a = 0$ (MAD00). The two-dimensional distributions of several physical quantities of MAD98 are shown in fig. S2. Specifically, the red line in each panel of fig. S2 denotes the boundary of the BZ-jet.

To calculate the radiation of the jet, we need to quantify the electron distribution, including both the thermal and nonthermal components. Because our MHD simulation is for a single fluid, we calculate the temperature of thermal electrons following the approach presented in (14), which is a function of the plasma β (the ratio between gas and magnetic pressure) and two other free parameters $R_{low}$ and $R_{high}$ (the "Determination of distributions of thermal and nonthermal electrons" section in Materials and Methods). A large fraction of radiation of the jet is believed to originate from nonthermal electrons, but the specific mechanism of electron acceleration has been a mystery. Here, we assume that the mechanism is magnetic reconnection. Numerical simulations have found that MADs are subject to magnetic flux eruptions that occur when a magnetic flux bundle containing strong vertical fields escapes from the black hole's magnetosphere and propagates radially outward into the disk (4, 15–19). These events must strongly perturb the vertical magnetic field lines and can in principle trigger magnetic reconnection in the jet. We have conducted several quantitative analyses and confirmed this scenario (in the "Physical origin of magnetic reconnection in the jet" section in Materials and Methods).

In light of the particle acceleration by magnetic reconnection (20–24), we assume that the accelerated electrons follow a power-law energy distribution. Details on how we determine the minimum Lorentz factor ($\gamma_{min}$) and power-law index ($p$) of the electron energy distribution are given in the "Determination of distributions of thermal and nonthermal electrons" section in Materials and Methods. For our problem, one of the most important parameters is the amount of nonthermal electrons and their spatial distribution. In the present work, we adopt the description proposed by Petersen *et al.* (*25*), which assumes that acceleration rate is proportional to $(\mathbf{J}/J_0)^2$, where $\mathbf{J}$ is the current density and $J_0$ is a characteristic current density characterizing the property of the background plasma. The physical motivation of such a prescription is based on particle-in-cell (PIC) simulations of particle acceleration (*25*). On the other hand, we would like to emphasize that this is still a simplified prescription to particle acceleration by reconnection because it relies on volumetric current rather than the current in the reconnecting current sheets. Radiative cooling is also considered when we try to obtain the number density of nonthermal electrons (the "Determination of distributions of thermal and nonthermal electrons" section in Materials and Methods). This description distinguishes our model from others. For instance, in (*26*), the number density of nonthermal electrons is roughly a constant percentage of the number density of thermal electrons, with mainly the spectral index of nonthermal electrons changing from place to place, depending on the local values of parameters β and σ.

The three panels in Fig. 1 show the two-dimensional distribution of $\mathbf{J}$, the ratio of the number density of nonthermal (power-law) electrons and total electrons $N_{pl}/N_{tot}$, and the number density of nonthermal electrons $N_{pl}$. We can see that $N_{pl}/N_{tot}$ reaches its largest values in a layer around the boundary of the BZ-jet. Note that the largest value of $N_{pl}/N_{tot}$ is close to but smaller than one. We find that most of the jet radiation comes from the part of the above-mentioned layer inside of the BZ-jet, and the narrowness of the radiation layer is one of the reasons for the limb-brightening feature we will show later. The number density of nonthermal electrons is largest in the equatorial plane of the accretion flow because the gas density there is much larger than in the jet. This does not imply that the radio emission from the accretion flow will dominate the jet because the magnetic field in the accretion flow is weaker than in the jet (fig. S2).

We use the general relativistic radiative transfer code IPOLE (*27*) to calculate the predicted images of the jet ("Calculation of radiative transfer in the jet" section in Materials and Methods). We tested 50 snapshots from the simulation data of MAD98 and found that our results are systematically and substantially better than previous works, in the sense that the jet is longer, its width is more consistent with observations, and its limb brightening is much more evident. Figure 2 compares the representative model images at 86 and 43 GHz, demonstrating that only the "current density" model successfully reproduces the opening angle of the jet and the elongated structure up to distance ∼2 mas from the core, while a model only considering thermal electrons fails. The calculated image terminates at ∼2 mas because it reaches the outer boundary of our simulation domain. For comparison, the best result obtained in previous work is only ∼1 mas (*26*), similar to our "thermal-only" model. Moreover, the long-standing puzzle of the observed limb-brightening feature of the jet is clearly reproduced, as shown by Fig. 3 ("Calculation of radiative transfer in the jet" section in Materials and Methods). By contrast, models that only consider thermal electrons or homogeneous distributions of nonthermal electrons produce very weak limb-brightening features (*26*, *28*). In addition, we have examined a test model in which $N_{pl}/N_{tot}$ is fixed to be 0.5, while all other model parameters remain the same with our fiducial model. The jet images predicted by this model have been calculated and are shown in fig. S6. Comparing the (flux contour) results with those presented in Fig. 2, we can see that this model is more similar to the thermal-only model rather than the current-density model because the limb-brightening feature is only present at large distance, and the jet is very short. The total flux densities predicted by our model at 86 and 43 GHz are 1.24 and 2.0 jansky (Jy), in comparison with the observed ranges of 0.95 to 1.59 Jy and 0.99 to 1.33 Jy at these two frequencies, of which 58 and 64% originates from nonthermal electrons. Because the plasma density in our model is normalized to reproduce the flux density at 230 GHz, our model correctly reproduces the flux densities at 230 and 86 GHz but moderately overpredicts that at 43 GHz. This is likely because our model is still too simplified in some aspects.

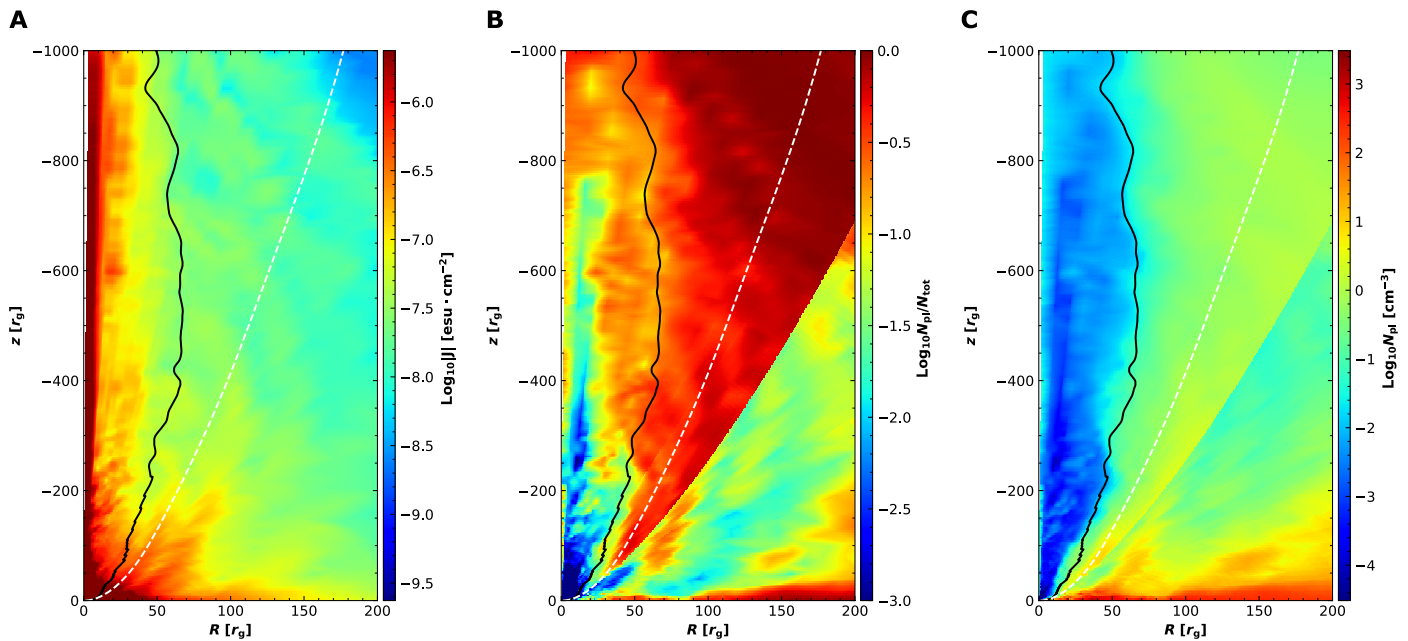

**Fig. 1. The φ-averaged two-dimensional spatial distribution of some quantities of MAD98 at a simulation time of *t* = 27,400.** From left to right, the colors denote the logarithm of (**A**) the three-current density **J**, (**B**) the ratio of number density of nonthermal and total electrons $N_{pl}/N_{tot}$, and (**C**) the number density of nonthermal (power-law) electrons $N_{pl}$. In each panel the black solid line denotes σ = 5, while the white dashed line denotes the φ-averaged boundary of the BZ-jet. In the middle panel, there is a sharp feature to the right of the white dashed line. This is due to the different scaling of $r_z$ in Eq. 6 adopted in our model. Its presence will not affect our results.

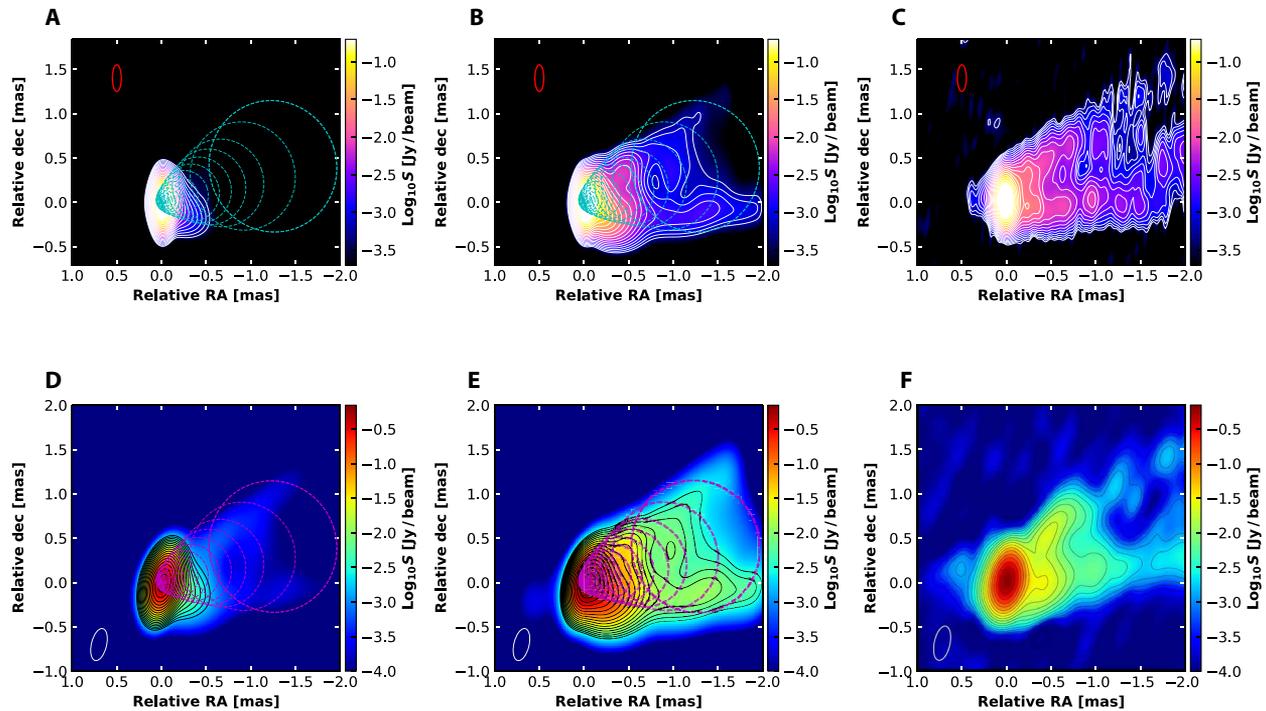

**Fig. 2. Comparison of images predicted by models and observations.** (**A** to **C**) Images at 86 GHz. (**D** to **F**) Images at 43 GHz. The left, middle, and right panels are the images predicted by (A and D) the thermal-only model, (B and E) the fiducial current-density model, and (C and F) the observed images, respectively. The observational data are taken from (*64*) (86 GHz) and (*65*) (43 GHz). The model images are calculated by convolving the simulations with a beam of 0.3 × 0.1 mas (top) and 0.37 × 0.17 mas (bottom) at −13.3° to mimic the limited resolution of the telescopes used in the observations. The dynamical ranges of the predicted and observed images are the same. See the "Calculation of radiative transfer in the jet" section in Materials and Methods for details for calculating the images. The dotted circles denote the boundary of the BZ-jet.

To quantify the jet morphology, Fig. 4 illustrates the variation of the jet width as a function of projected distance from the core as predicted by different models ("Calculation of the jet width" section in Materials and Methods) and compares them with observations. The jet width predicted by the MAD98 model agrees reasonably well with observations throughout the entire projected jet extent from 0.1 to 2 mas. This improvement compared to previous works, along with the better match of the length and limb-brightening of the jet, is attributed to our more physical model for the nonthermal electron population. That the jet width is smaller than the boundary between a BZ-jet and disk-jet (cyan line in Fig. 4) indicates that the observed jet corresponds to the former instead of the latter.

Our fiducial model also satisfies the constraints based on the velocity field, jet power, and polarization (text S2). In text S3, we investigate the effects of black hole spin and the accretion mode on the predicted jet image. We find that a MAD and a rapid spin are favored to explain the observations. In some jet models, the emission is dominated by electron-positron pairs within the jet, even close to the horizon scale. The possibility of giving constraint on this mechanism based on our model is discussed in text S4.

Our results bridge the dynamical model of jet formation and observations, demonstrating the viability of the BZ-jet model and magnetic reconnection as the mechanism for electron acceleration in jets. The models will be sharpened in the future by incorporating more detailed particle acceleration by reconnection, obtaining more elaborate electron energy distribution in the jet, and by comparison with observations of higher resolution.

## MATERIALS AND METHODS

### Three-dimensional GRMHD numerical simulation of a hot accretion flow

We have performed numerical simulations in three-dimension by solving the equations of ideal MHD describing the evolution of the accretion flow around a black hole in the Kerr metrics using the GRMHD code Athena++ (*29*, *30*). The code uses a finite-volume Godunov scheme to ensure total energy conservation, with the flux of conserved quantities obtained by solving the Riemann problem at each interface. The commonly used HLLE Riemann solvers are used in our simulations (*31*). The staggered-mesh constrained transport method is applied to satisfy the divergence-free constraint to prevent spurious production of magnetic monopoles. We use the piecewise linear method for spatial reconstruction. It will be interesting to examine the effects of adopting the piecewise parabolic method in the future (*32*).

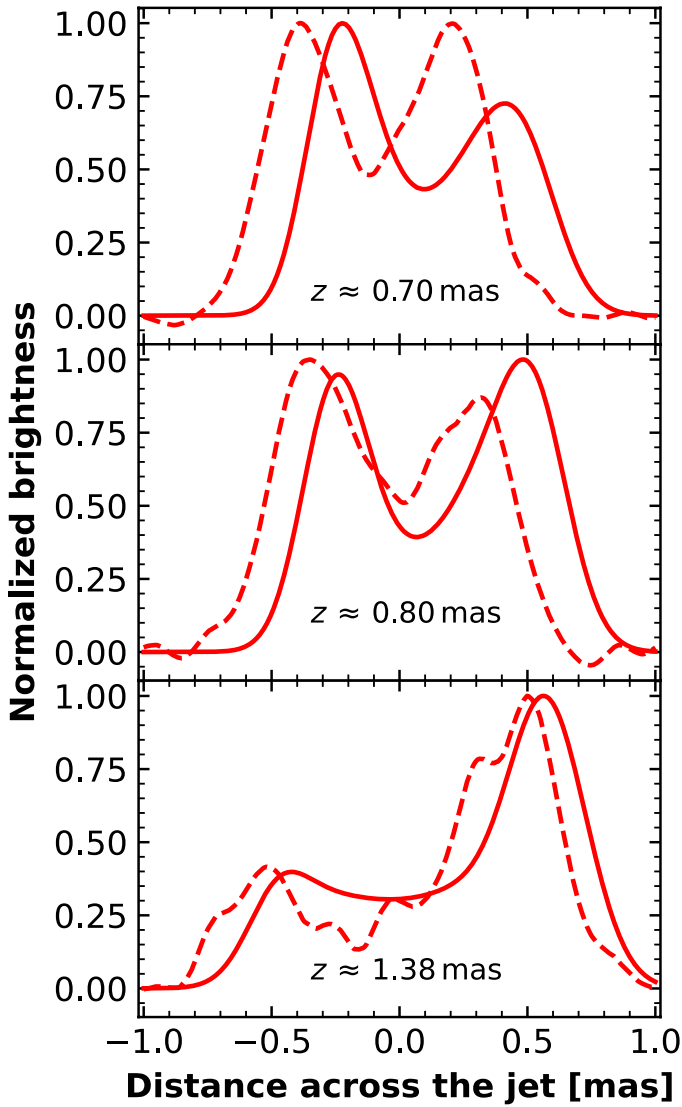

**Fig. 3. The limb-brightening predicted by the fiducial model (solid lines) and its comparison with observations (dashed lines) at three distances from the core.** The observational data are based on Global mm-VLBI Array (GMVA) observations at 86 GHz (64). As the flux fluctuates with time for both the models and observations, we have normalized the predicted flux to the observed value.

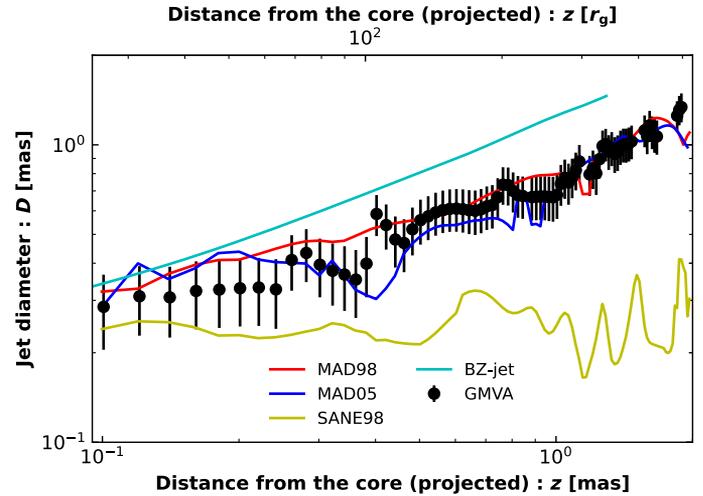

**Fig. 4. Comparison between the predicted and observed jet width as a function of the projected distance from the core.** The observational data are based on stacked GMVA observations at 86 GHz (64). The cyan line corresponds to the boundary of the BZ-jet. The fiducial MAD98 model well reproduces the observed jet width throughout the projected jet extent from 0.1 to 2 mas but underpredicts the jet width within 0.1 mas recently measured by Lu et al. (40). This is likely because the resolution of our simulations is not high enough ("Convergence of numerical resolution" section in Materials and Methods).

We perform all our simulations in Kerr-Schild (horizon penetrating) coordinates ($t$, $r$, θ, φ). The radius of the black hole horizon is $r_{\mathrm{H}} = (1 + \sqrt{1 - a^2})r_{\mathrm{g}}$, with $r_{\mathrm{g}} = \frac{GM_{\mathrm{BH}}}{c^2}$ the gravitational radius and $a$ the spin parameter of the black hole. The determinant of the metric $g \equiv \det(g_{\mu\nu}) = -\Sigma^2 \sin^2\theta$, where $\Sigma = r^2 + a^2\cos^2\theta$ (33). We use ρ to denote the comoving rest-mass density, and $u^\mu$ as the component of the coordinate-frame four velocity. The equation of state of the gas is $u = p_{\mathrm{gas}}/(\Gamma - 1)$, where the $p_{\mathrm{gas}}$ is the gas pressure of the comoving mass, $u$ is the internal energy of the gas, and Γ is the adiabatic index. We set Γ = 4/3 in our simulation and use unit $r_{\mathrm{g}}/c$ to measure the time. The unit we adopt is Heaviside-Lorentz, in which both the light speed and gravity constant are set to be unity, and the sign convention of the metric is (−, +, +, +). The metric is stationary, and the self-gravity of the accretion flow is ignored in our simulations.

There are two accretion modes, SANE and MAD, depending on the magnetization of the accretion flow (5, 34, 35). We have simulated four models, namely, MAD98, MAD05, MAD00, and SANE98. They denote the MAD accretion flow around a black hole of spin $a$ = 0.98, 0.5, and 0.0, and SANE accretion flow around a black hole of spin $a$ = 0.98. In all models, the simulation starts with a rotating torus around a black hole. The torus is in a hydrostatic equilibrium state as described by Fishbone and Moncrief (36). The inner edge of the torus is at $r$ = 40.5$r_{\mathrm{g}}$, and the radius of pressure maximum is at $r$ = 80$r_{\mathrm{g}}$. We have also added a poloidal magnetic field threading the torus in the way described by Penna et al. (37). For MAD and SANE, we use two different initial magnetic field configurations. For MAD, we set one poloidal loop threading the whole torus. This leads to rapid accretion and accumulation of a large amount of magnetic flux near the black hole, which eventually impedes the accretion of mass. For a given mass accretion rate, the magnetic flux quickly saturates at a maximum value on the black hole, reaching the MAD state. For SANE, we initially set up a seed field consisting of multiple poloidal loops of magnetic field with changing polarity. This structure facilitates magnetic reconnection and prevents the accumulation of magnetic flux, such that the magnetic field always remains weak, making the accretion flow stay in the SANE mode (34, 35).

To determine the specific magnetic field of the torus, three parameters are required, namely, $r_{\mathrm{start}}$, $r_{\mathrm{end}}$ and $\lambda_B$. The first two represent the inner and outer edges of the magnetized area, while the last one controls the size of the poloidal loops or equivalently the number of the loops. We use the gas-to-magnetic pressure ratio β ≡ $p_{\mathrm{gas}}/p_{\mathrm{mag}}$ to normalize the magnetic field. For each loop, it has a minimum value in the equatorial plane. It peaks at the edge of the loop and then drops to the center of the loop. For MAD, we set $r_{\mathrm{start}}$ = 25 $r_{\mathrm{g}}$, $r_{\mathrm{end}}$ = 810 $r_{\mathrm{g}}$, $\lambda_B$ = 25 (it just has one loop), and $\beta_{\mathrm{min}}$ = 0.1. For SANE, we set $r_{start} = 25 r_g$, $r_{end} = 550 r_g$, $\lambda_B = 2.5$ (it has eight loops), and $\beta_{min} = 0.1$ ($a = 0.98$). We have performed test simulations with higher values of $\beta_{min} = 1, 10, 100$ and found that the results remain very similar.

We use a static mesh refinement grid, which includes a root grid and additional refined grids in the jet region of interest to the present work. For MAD98, the root grid is $88 \times 32 \times 16$ cells in the radial, polar, and azimuthal directions. We adopt logarithmic spacing, with the ratio $\frac{r_{i+1}}{r_i}$ being 1.0827 in the $r$ direction. For SANE98, MAD05, and MAD00, the root grid is $110 \times 32 \times 16$ cells in the radial, polar, and azimuthal directions, and the ratio $\frac{r_{i+1}}{r_i}$ is 1.0667 in the $r$ direction. For MAD98 (SANE98, MAD05, and MAD00), their inner and outer edges are located at 1.1 $r_g$(1$r_g$) and 1200 $r_g$, respectively. The grid in the polar and azimuthal directions are uniform. Because Kerr-Schild coordinates are used, the simulated inner edge is within the black hole horizon.

We use different static mesh refinement for different models. Tables S1 and S2 give the details of the four models. For a refinement region, each additional level means that the grid of this area is refined by a factor of 2 for all directions based on the previous level grid. Note that in ATHENA++, a higher level SMR block has not to be contained within a lower level SMR block. Level 4 is added to improve the resolution of the most important jet region for our modeling. However, to save computation resources, we only add this level in the forward jet region but not in the backward jet region. The last entry in level 3 is added to emphasize this special grid setting. The final grid level for MAD98 is shown in fig. S1. To prevent unphysical (purely numerical) causal contact, our setup ensures that there are at least four cells within the event horizon in all directions for the four models. The final effective resolution upon which our fiducial model is based on $1408 \times 512 \times 256$ for MAD98 and $880 \times 256 \times 128$ for SANE98, MAD05, and MAD00. This effective resolution is achieved within a range of radii, which is most important for our modeling.

Within the BZ-jet region, we have 32 grids at $z \sim 700 r_g$ and 44 grids at $z \sim 100 r_g$ in the $\theta$-direction for MAD98. The number of grids within the BZ-jet region at $z \sim 100 r_g$ for SANE98, MAD05, and MAD00 are 15, 24, and 24, respectively.

We use the outflow boundary conditions at the inner and outer boundaries. For the $\theta$ and $\varphi$ direction, we use the polar axis reflective boundary condition and periodic boundary conditions respectively. Because vacuum cannot be handled in numerical simulations, the density and gas pressure floors must be imposed: $\rho_{min} = \max(10^{-2}r^{-1.5}, 10^{-6})$, $p_{gas,min} = \max(10^{-2}r^{-2.5}, 10^{-8})$. We also enforce $\sigma < 100$ and $\gamma < 50$. Here, $\sigma = 2p_{mag}/\rho$ is the magnetization parameter, and $\gamma \equiv \alpha u^t$, with the lapse $\alpha \equiv (-g^{tt})^{-0.5}$. The former is an additional limitation of the density and gas pressure floors; the latter limits the velocity to prevent the Lorentz factor of the normal frame from becoming too large (38).

We run simulations up to $t_f = 40,000$ for MAD98, MAD05, MAD00, and SANE98. They correspond to 8.9 orbital periods of the accretion flow at the pressure maximum. The "inflow equilibrium" reached about 80 $r_g$ for the three MAD models and 30 $r_g$ for SANE98. Figure S2 shows the distribution in the $x$-$z$ plane of several physical quantities for MAD98, while fig. S3 shows the distribution in the $x$-$y$ plane at $z = 600 r_g$ of physical quantities for this model. In all models except MAD00, a BZ-jet is successfully produced. The outer boundary of a BZ-jet is defined as the time-averaged surface consisting of magnetic field lines anchored at the event horizon of the rotating black hole with $\theta = 90°$, which also serves as the boundary between the BZ-jet and disk-jet. This is shown by the red lines in each panel of fig. S2. In the present work, we do not perform a detailed analysis of the BZ-jet and disk-jet. Analysis of the BZ-jet can be found in (3, 4), while the analysis of the disk-jet in the case of SANE can be found in (10).

We believe that the jets in our simulations have reached a steady state. Detailed virtual particle trajectory analysis indicates that both jet and wind close to the rotation axis are produced at small radii the accretion flow (10). In addition, the velocity of the jet material is several orders of magnitude higher than that of the accretion flow; thus, $z/v_z$ in the jet is much smaller than the accretion time scale of the accretion flow. So once the innermost region of the accretion flow has reached steady state, as in our case, the jet should also very quickly reach its steady state.

### Convergence of numerical resolution

We first examine the convergence of MHD turbulence driven by magneto-rotational instability (MRI). We follow (39) to calculate $Q_\theta$ and $Q_\phi$ based on our simulation data

$$Q_\theta = \frac{\lambda_{MRI}}{dx^\theta} = \frac{2\pi|v_{\theta,A}|}{\Omega dx^\theta} \tag{1}$$

$$Q_\phi = \frac{\lambda_{MRI}}{dx^\phi} = \frac{2\pi|v_{\phi,A}|}{\Omega dx^\phi} \tag{2}$$

Here $V_{\theta,A}$ and $V_{\phi,A}$ are the $\theta$- and $\phi$-directed Alfven speeds, and $\Omega$ is the angular velocity of fluid. The physical meaning of these two quantities is the number of grid in one fastest growing wavelength of MRI in the $\theta$- and $\phi$-directions, respectively. According to Hawley et al. (39), if $Q_\theta > 10$ and $Q_\phi > 20$, the resolution of the simulation will be high enough to resolve MRI and give quantitatively converged results for the nonlinear MHD turbulence in the accretion flow. We have calculated these two parameters and found that both are larger than 50 in almost all regions of our simulation domain.

We also need to examine the convergence of current density calculations, especially because in our MHD simulations, dissipation is not explicitly modeled. This issue has been discussed in (25). They show that even if the current density does not converge with resolution, the total dissipation rate does.

Last, to examine the convergence of our main results, we have conducted a simulation of MAD98 model with a lower resolution of $352 \times 128 \times 64$. This resolution is lower by a factor of 4 in each direction than our fiducial MAD98 model. We then have examined the convergence in the following two ways. One is that we have compared both the time and $\phi$-averaged "basic physical quantities" (i.e., density, temperature, and plasma $\beta$) in the two models. The results are shown in fig. S4, indicating reasonable convergence. In addition, using the low-resolution simulation data, we have also produced the jet image at 86 GHz and examined the limb brightening and the jet width. These results are shown in fig. S5. Comparing the results shown in fig. S5 with corresponding results predicted by the high-resolution model, we again find satisfactory resolution convergence. However, note that the limb-brightening feature predicted by the low-resolution model is not as good as that predicted by the high-resolution fiducial model, indicating the role of adopting higher resolution.

In addition to limb brightening, the predicted jet width may also be related with the simulation resolution. We find that our fiducial MAD98 model underpredicts the jet width at $z < 0.1$ mas from the black hole most recently measured by Lu *et al.* (*40*). We think the main reason is that we fail to resolve the limb brightening at this region of the jet, which is likely because our simulation resolution at $r \leq 30r_g$ is still not high enough (refer to table S1). It will be interesting to test this possibility by increasing the simulation resolution in the future, although the calculation will be very expensive.

### Determination of distributions of thermal and nonthermal electrons

The temperature of thermal electrons in the simulation is determined using the following formula (*14*)

$$\frac{T_p}{T_e} = R_{low}\frac{1}{1+\beta^2} + R_{high}\frac{\beta^2}{1+\beta^2} \quad (3)$$

Here $T_p$ and $T_e$ are the proton and electron temperature, and $\beta$ is the ratio between the gas and magnetic pressure. We have tested different values of $R_{low}$ and $R_{high}$ and found that $R_{low} = 1$ and $R_{high} = 80$ provide the best result. These two values are also commonly used in the literature. We note that our value of $R_{low}$ is different from that obtained in Event Horizon Telescope Collaboration (EHTC) (*12*), where its value is required to be 10 for the MAD model with a high black hole spin. One possible reason for the discrepancy is that their value is based on an assumption that the observed low polarization of the nuclear region of M87 is due to Faraday rotation internal to the emission region. This assumption is not adopted in our model. In addition, the EHT work focuses on the accretion flow while our work focuses on the jet. The properties of the plasma in the accretion flow and jet, such as the plasma sigma, are notably different thus the electron heating may be different.

The nonthermal electrons accelerated by magnetic reconnection are assumed to be described by the following power-law form

$$\frac{dn_{pl}}{d\gamma} = N_{pl}(p-1)\gamma^{-p}, \gamma_{max} > \gamma > \gamma_{min} \quad (4)$$

In the literature, the value of power-law index $p$ is often treated as a constant free parameter. In our work, however, we use the most recent results obtained in (*24*). In this work, they present a model for determining the value of $p$, with the processes of particle injection, acceleration, and escape included. They use the results of a series of first-principle fully kinetic simulations to calibrate a couple of model coefficients. This model not only can successfully reproduce the simulation results but also can predict the power-law index $p$. The value of $p$ is described by

$$p = \frac{1}{\sigma_x + 0.2[1+\tanh(b_g)]} + 0.04\tanh(b_g)\sigma_x + 1.7b_g + 2.1 \quad (5)$$

Here $b_g$ is used to describe the guide field $B_g = b_g B_0$, $B_0$ is the reconnection magnetic field, $\sigma_x = B_0^2/w$, with $w = (\rho + P_{gas} + u_{gas} + B_0^2)$ is the enthalpy density. In our work, we consider the magnetic field from our GRMHD simulation as the total magnetic field, so $B = \sqrt{B_0^2 + B_g^2}$. We set $b_g = 0.5$. We have performed several tests and found that our results are not sensitive to its value. The distribution of the value of $p$ is shown by fig. S7.

PIC simulations of particle acceleration by reconnection find that the exact value of $\gamma_{min}$ depends on many parameters such as $\beta$ and the guide field and is still poorly determined (*41*). Different ways are adopted in the literature to set the value of $\gamma_{min}$. Özel *et al.* (*42*) and Yuan *et al.* (*43*) require that the number density of nonthermal electrons must be equal to that of thermal electrons at $\gamma_{min}$. Dexter (*44*) treats it as a free parameter, while Chatterjee (*45*) ties $\gamma_{min}$ to the peak of the thermal distribution, namely, $\gamma_{min} = u_{th}/m_ec^2 + 1 \approx 3kT_e/m_ec^2 + 1$ for relativistic electron temperature. Here, $u_{th}$ is the energy density of thermal electrons. In our work, we set a higher value, $\gamma_{min} = 10kT_e/m_ec^2 + 1$, at which particle energy is weakly suprathermal. The value of $\gamma_{max}$ is not important for the calculation of radiation if the value of $p$ is not too small as in our case. We have tested two values of $\gamma_{max}$, $10^5$ and $10^6$, and the differences of results are found to be negligible.

For our problem, the most important quantity to determine is the amount of nonthermal electrons and their spatial distribution. Many different ways have been adopted to specify this parameter in the literature. In some works, its value is determined by assuming that the energy density of nonthermal electrons is a constant fraction of the energy density of thermal electrons (*42, 43, 46, 47*) or magnetic field (*44, 48*). The dependence of acceleration efficiency in magnetic reconnection on the plasma $\beta$ and magnetization parameter $\sigma$ is investigated by PIC simulations in (*22, 24*), and the result is adopted in (*45*) and (*49*) in their studies of flares from accretion flows.

In our work, we assume that the electrons are accelerated by magnetic reconnection. Following (*25*), we assume that the acceleration rate is proportional to the square of the local three-current density measured in the frame of the plasma, $\mathbf{J}^2$. We found that the radiative time scale of accelerated electrons is different at different distances in the jet and can be shorter than the local dynamical time scale; therefore, radiative cooling should be included when we try to obtain a steady energy distribution of nonthermal electrons. Given the above analysis, the steady number density of nonthermal electrons $N_{pl}$ is determined by solving the following equation

$$\eta\frac{\nu_A}{r_z}(N_{tot} - N_{pl})\frac{\mathbf{J}^2}{J_0^2} = \frac{N_{pl}}{\tau_{cool}} \quad (6)$$

Here, $N_{tot}$ is the total number density of electrons, including thermal and power-law electrons, which is taken directly from our simulations; $\eta$ is a dimensionless parameter that controls the efficiency with which currents accelerate electrons into the power-law distributions, with its value being given in table S3 for various models; and $\nu_A$ is the Alfven speed, which determines the magnitude of the reconnection inflow speed (*20, 21*). The difference between the reconnection inflow speed and $\nu_A$ is absorbed in $\eta$. $r_z$ denotes the typical length scale of the reconnection region. Within the jet region, the reconnection is likely driven by the strong perturbation of the magnetic field lines caused by the magnetic eruption in the accretion flow (see the "Physical origin of magnetic reconnection in the jet" section in Materials and Methods for the discussion), so $r_z$ should be the scale of jet width. Motivated by the observed jet width as a function of jet distance (refer to Fig. 4), we set $r_z = z^{1/3}$. Outside of the jet region, we simply set $r_z = r$. The boundary between these two scalings of $r_z$ should be beyond the BZ-jet boundary because the perturbation to the field lines caused by the magnetic eruption originates from the innermost region of the accretion flow, and it should propagate mainly in the polar region along the field lines. In our calculation, we therefore choose a line described by $R = 2 + 2 \times z^{0.7}$ as the boundary of the two different scalings. This boundary well mimics the shape of a magnetic field line beyond the BZ-jet boundary. The discontinuity at the boundary produces the sharp feature of $N_{pl}/N_{tot}$ presented in Fig. 1B. We have performed tests by adopting different boundaries for different scalings and by adopting the same scaling in the whole region. We have confirmed that the discontinuity here will not affect our main results. Especially, the limb brightening of the jet remains unchanged. $\tau_{cool}$ is the local radiative cooling time scale of nonthermal electrons. In general, it depends on electron Lorentz factor $\gamma$. In the present work, given that the value of power-law index determined by Eq. 5 is usually large, $\sim 4 - 6$ (refer to fig. S7), we adopt a simplification by only considering electrons with $\gamma_{min}$ when we estimate $\tau_{cool}$. This simplification is similar to that adopted in "model D" in (25). $\mathbf{J}$ is the local three-current density, which is calculated following (50)

$$J^i = \partial_j F^{ij} + \Gamma^i_{jk} F^{ij} \tag{7}$$

where $F^{\mu\nu}$ is the electromagnetic tensor, $\Gamma^\lambda_{\alpha\beta}$ is the Christoffel symbols, and $i$ equals to $t$, $r$, $\theta$, and $\phi$. The characteristic three-current density $J_0^2 \equiv c^2 P / r^2$, with $P$ and $r$ being the gas pressure and spherical radius, respectively (25).

## The physical origin of magnetic reconnection in the jet

Some works have suggested that the jet is subject to current-driven kink instability, which can disrupt magnetic field lines and trigger the magnetic reconnection (51–53). According to the Kruskal-Shafranov (KS) criterion, cylindrical configurations in which the toroidal component of the magnetic field dominates are unstable to the $m = 1$ kink mode. The classical KS criterion is extended when the relativistic field line rotation is taken into account (54). Jets are unstable only if both the KS criterion and the condition $|B_\phi/B_p| > R\Omega/c$ are satisfied. Here, $R$ is the cylindrical radius of jet, $\Omega$ is the angular velocity of the field lines, and $c$ is the speed of light. We have evaluated the values of $B_\phi$ and $B_p$ in the comoving fluid frame and found that these two values are overall comparable to each other, although strong spatial fluctuation exists; thus, the jet is not subject to kink instability. This is especially true in the small distance in the jet where the poloidal component of the magnetic field is dominant. However, in that case, nonaxisymmetric features of the current density distribution are found to be as strong as in other distances, as we will show later.

We then consider another possibility; that is, the reconnection in the jet is caused by magnetic eruption in MAD (4, 15–19). MHD numerical simulations of MAD have found that the poloidal field will be carried inward by the accretion flow and accumulate on the black hole horizon. When the field pressure overcomes the dynamic pressure of the accreting gas, the magnetic field quasi-periodically erupts. The eruption occurs in the form of low-density, magnetically dominated medium fountained outward from the black hole. This will produce a strong perturbation to the field lines, and the perturbation should propagate vertically outward and trigger magnetic reconnection in the jet. In addition to this mechanism, the high-resolution simulation by Ripperda et al. (18) indicates that the eruption is accompanied by reconnection in the accretion flow near the horizon, which transforms the horizontal field to a vertical field that is then ejected outward vertically and also triggers reconnection in the jet region.

To examine this possibility, we have performed several tests. Because the eruption is found to be nonaxisymmetric, the distribution of current density in the $x - y$ plane of the jet caused by this mechanism should be nonaxisymmetric. Using the MAD98 simulation data, we have calculated the distribution of current density and did find a nonaxisymmetric current density distribution, as shown in fig. S8. Next, because eruption events occur in MAD but not in SANE, we should expect that the nonaxisymmetric feature should be much stronger in MAD than in SANE. To examine this point, we have performed a power spectrum analysis of the magnitude of the current density $\mathbf{J}$ by performing the volume-averaged Fourier transform following the approach presented in (55)

$$f(m, k) = \frac{1}{V_{cl}} \iiint_{V_{cl}} |\mathbf{J}| \, e^{i(m\phi + kz)} r dr d\phi dz \tag{8}$$

where $|\mathbf{J}|$ is the magnitude of $\mathbf{J}$ and $V_{cl}$ is the cylindrical volume that encloses $\mathbf{J}$

$$V_{cl} = \int_{r_a}^{r_b} \int_0^{2\pi} \int_{z_a}^{z_b} r dr d\phi dz \tag{9}$$

The quantity $f(m, k)$ is a function of the azimuthal mode number $m$ and the axial wave number $k = 2\pi/\lambda$ ($\lambda$ is the characteristic wavelength). The power spectrum is calculated as

$$|f(m, k)|^2 = \{\text{Re}[f(m, k)]\}^2 + \{\text{Im}[f(m, k)]\}^2 \tag{10}$$

$$\text{Re}[f(m, k)] = \frac{1}{V_{cl}} \iiint_{V_{cl}} |\mathbf{J}| \cos(m\phi + kz) r dr d\phi dz \tag{11}$$

$$\text{Im}[f(m, k)] = \frac{1}{V_{cl}} \iiint_{V_{cl}} |\mathbf{J}| \sin(m\phi + kz) r dr d\phi dz \tag{12}$$

The typical wavelength can be estimated by the product of the jet speed and eruption period, which are roughly speed of light and 1000 $r_g/c$, respectively (refer to fig. S9 for the eruption period). This means that the wave number $\kappa = 2\pi/\lambda \ll 1$; thus, we simply set $k = 0$ in our calculation. The time evolution of the power spectrum for $m = 1$ and $m = 0$ for MAD98 and SANE98 models at $z \sim 120 r_g$ and $z \sim 600 r_g$ in the jet are shown in the two panels of fig. S9. We can see from the figure that, in both panels, the $m = 1$ mode power (normalized by the $m = 0$ mode power) in the case of MAD98 is much larger than SANE98, consistent with our expectation. We note that although at small $z$ ($\sim 120 r_g$) where the poloidal magnetic field component is much larger than the toroidal component, thus the kink instability is expected to be absent, the result still remains the same, indicating again that kink instability is not the physical mechanism of driving reconnection. In addition, the typical time scale of the variability of mode power is roughly consistent with the variability time scale of the magnetic flux threading the horizon (56), again consistent with the magnetic eruption scenario. In the case of SANE98, we do not expect the existence of magnetic eruption, but we still find some power from fig. S9, although it is weak. We speculate that the power in this case arises from other mechanisms, such as some kinetic instabilities driven by the interaction between the jet and its ambient medium (*57*) or velocity shear close to the jet boundary (*58*).

### Calculation of radiative transfer in the jet

Because the plasma thermodynamics predicted by any energy conserving simulations is unreliable in highly magnetized region, in our calculation of radiation from jet, the emission originating from the highly magnetized region ($\sigma > \sigma_{cut}$) is excluded, as usual. The value of $\sigma_{cut}$ is highly uncertain, and different values are adopted in the literature (*26*, *28*, *59*). In many cases, a low value of $\sigma_{cut} = 1$ is adopted in the study of radiation from SANE. However, because the jet region is highly magnetized, higher values are often adopted for jets. For example, a fiducial value of $\sigma_{cut} = 25$ is adopted in (*28*), while $3 \leq \sigma_{cut} \leq 5$ is favored in (*26*). Here, we have tested four different values $\sigma_{cut} = 1, 3, 5, 6$, and 10. We find that the best results are achieved when $\sigma_{cut} = 5$ for our fiducial MAD98 model, and thus, we choose this value. In text S1, we have studied the effects of choosing different boundaries of the radiation region in the jet and found that our results are not very sensitive to the choice of the boundary. For the choice of a notably higher value of $\sigma_{cut}$, the predicted total flux density at 86 GHz would be too high compared to observations (we use the observed 230 GHz flux density to normalize our accretion rate). In addition, the predicted jet width would be too narrow because the radiation would concentrate more toward the jet axis. The results would go to the other extreme if a notably lower $\sigma_{cut}$ were adopted.

Because our GRMHD simulation is density scale-free, we need to normalize the density and magnetic field of the simulations. We do this by choosing a density so that the accretion can reproduce the observed radio flux density of 0.5 Jy at 230 GHz obtained by EHT for M87 in the field of view of 100 μas (*60*). The mass of the black hole of M87 is adopted to be $6.5 \times 10^9 M_\odot$ (mass of the Sun) (*61*). The obtained mass accretion rates for difference models are listed in table S3. We use IPOLE, which is a public ray-tracing code for covariant, polarized radiative transport, to calculate the image of the jet corresponding to various models (*27*). Note that in our calculations of images, the contributions of all plasma in the simulation domain, including the accretion flow and wind launched from the accretion flow, have been included.

For MAD98, we have randomly chosen 50 different snapshots from the simulation data after the simulation has achieved steady state and calculated their corresponding images. The viewing angle of M87 is 163°, and the distance to the black hole is 16.8 Mpc. The field of view of our calculated image is [−2000, 2000] μas both in $x$ and $y$ directions, and the resolution of the image is $500 \times 500$. After completing the radiation transfer calculation, we rotate the output image by 108° counterclockwise to make the orientation of the jet consistent with the observed position angle of about 288° (*62*). We have added in the images some flux density contours in the following way. Taking the 86 GHz image as an example, denoting the peak flux density of the predicted image as $S_{predict}$ Jy per beam, the value of each contour in the image is $[−0.47, 0.47, 0.47 \times \sqrt{2}, 0.47 \times 2, \ldots] \times 10^{-3} S_{predict} / S_{obs}$ Jy per beam, i.e., increasing by factor of $\sqrt{2}$ for two adjacent contours until the peak flux density is reached, as in the observed image. Here, $S_{obs}$ (=0.560 Jy per beam) and $0.47 \times 10^{-3}$ Jy per beam are the peak and the lowest flux density of the observed image, respectively. In this way, we ensure to have the same number of contours in both the theoretical and observed images, so that the dynamical ranges of flux density in the theoretical and observed images are the same.

### Calculation of the jet width

We calculate the width of the jet following the same approach of defining the jet width in the analysis of observations (*63*). That is, based on the calculated jet image, we slice the jet in the direction perpendicular to the jet axis and fit the flux-density profile with one or three Gaussians. The jet width is then defined as the deconvolved full width at half maximum of the Gaussian profile or the distance between the two outermost Gaussians.

## REFERENCE AND NOTES

1. R. D. Blandford, R. L. Znajek, Electromagnetic extraction of energy from Kerr black holes. *Mon. Not. R. Astron. Soc.* **179**, 433–456 (1977).
2. J. F. Hawley, S. A. Balbus, The dynamical structure of nonradiative black hole accretion flows. *Astrophys. J.* **573**, 738–748 (2002).
3. A. Tchekhovskoy, R. Narayan, J. C. McKinney, Efficient generation of jets from magnetically arrested accretion on a rapidly spinning black hole. *Mon. Not. R. Astron. Soc.* **418**, L79–L83 (2011).
4. J. C. McKinney, A. Tchekhovskoy, R. D. Blandford, General relativistic magnetohydrodynamic simulations of magnetically choked accretion flows around black holes. *Mon. Not. R. Astron. Soc.* **423**, 3083–3117 (2012).
5. A. Sadowski, R. Narayan, R. Penna, Y. Zhu, Energy, momentum and mass outflows and feedback from thick accretion discs around rotating black holes. *Mon. Not. R. Astron. Soc.* **436**, 3856–3874 (2013).
6. R. D. Blandford, D. G. Payne, Hydromagnetic flows from accretion discs and the production of radio jets. *Mon. Not. R. Astron. Soc.* **199**, 883–903 (1982).
7. D. Lynden-Bell, On why discs generate magnetic towers and collimate jets. *Mon. Not. R. Astron. Soc.* **341**, 1360–1372 (2003).
8. Y. Kato, S. Mineshige, K. Shibata, Magnetohydrodynamic accretion flows: Formation of magnetic tower jet and subsequent quasi-steady state. *Astrophys. J.* **605**, 307–320 (2004).
9. J. F. Hawley, J. H. Krolik, Magnetically driven jets in the kerr metric. *Astrophys. J.* **641**, 103–116 (2006).
10. F. Yuan, Z. Gan, R. Narayan, A. Sadowski, D. Bu, X-N. Bai, Numerical Simulation of Hot Accretion Flows. III. Revisiting Wind Properties Using the Trajectory Approach. *Astrophys. J.* **804**, 101 (2015).
11. Materials and methods are available at the end of the main text where supplementary materials are available at the Science Advances website.
12. Event Horizon Telescope Collaboration, K. Akiyama, J. C. Algaba, A. Alberdi, W. Alef, R. Anantua, K. Asada, R. Azulay, A.-K. Baczko, D. Ball, M. Baloković, J. Barrett, B. A. Benson, D. Bintley, L. Blackburn, R. Blundell, W. Boland, K. L. Bouman, G. C. Bower, H. Boyce, M. Bremer, C. D. Brinkerink, R. Brissenden, S. Britzen, A. E. Broderick, D. Broguiere, T. Bronzwaer, D.-Y. Byun, J. E. Carlstrom, A. Chael, C.-K. Chan, S. Chatterjee, K. Chatterjee, M.-T. Chen, Y. Chen, P. M. Chesler, I. Cho, P. Christian, J. E. Conway, J. M. Cordes, T. M. Crawford, G. B. Crew, A. Cruz-Osorio, Y. Cui, J. Davelaar, M. De Laurentis, R. Deane, J. Dempsey, G. Desvignes, J. Dexter, S. S. Doeleman, R. P. Eatough, H. Falcke, J. Farah, V. L. Fish, E. Fomalont, H. A. Ford, R. Fraga-Encinas, P. Friberg, C. M. Fromm, A. Fuentes, P. Galison, C. F. Gammie, R. García, Z. Gelles, O. Gentaz, B. Georgiev, C. Goddi, R. Gold, J. L. Gómez, A. I. Gómez-Ruiz, M. Gu, M. Gurwell, K. Hada, D. Haggard, M. H. Hecht, R. Hesper, E. Himwich, L. C. Ho, P. Ho, M. Honma, C.-W. L. Huang, L. Huang, D. H. Hughes, S. Ikeda, M. Inoue, S. Issaoun, D. J. James, B. T. Jannuzi, M. Janssen, B. Jeter, W. Jiang, A. Jimenez-Rosales, M. D. Johnson, S. Jorstad, T. Jung, M. Karami, R. Karuppusamy, T. Kawashima, G. K. Keating, M. Kettenis, D.-J. Kim, J.-Y. Kim, J. Kim, J. Kim, M. Kino, J. Y. Koay, Y. Kofuji, P. M. Koch, S. Koyama, M. Kramer, C. Kramer, T. P. Krichbaum, C.-Y. Kuo, T. R. Lauer, S.-S. Lee, A. Levis, Y.-R. Li, Z. Li, M. Lindqvist, R. Lico, G. Lindahl, J. Liu, K. Liu, E. Liuzzo, W.-P. Lo, A. P. Lobanov, L. Loinard, C. Lonsdale, R.-S. Lu, N. R. MacDonald, J. Mao, N. Marchili, S. Markoff, D. P. Marrone, A. P. Marscher, I. Martí-Vidal, S. Matsushita, L. D. Matthews, L. Medeiros, K. M. Menten, I. Mizuno, Y. Mizuno, J. M. Moran, K. Moriyama, M. Moscibrodzka, C. Müller, G. Musoke, M. Mus, D. M. Alejandro, A. Nadolski, H. Nagai,

N. M. Nagar, M. Nakamura, R. Narayan, G. Narayanan, I. Natarajan, A. Nathanail, J. Neilsen, R. Neri, C. Ni, A. Noutsos, M. Nowak, H. Okino, H. Olivares, G. N. Ortiz-León, T. Oyama, F. Özel, D. C. M. Palumbo, J. Park, N. Patel, U.-L. Pen, D. W. Pesce, V. Piétu, R. Plambeck, A. PopStefanija, O. Porth, F. M. Pötzl, B. Prather, J. A. Preciado-López, D. Psaltis, H.-Y. Pu, V. Ramakrishnan, R. Rao, M. G. Rawlings, A. W. Raymond, L. Rezzolla, A. Ricarte, B. Ripperda, F. Roelofs, A. Rogers, E. Ros, M. Rose, A. Roshanineshat, H. Rottmann, A. L. Roy, C. Ruszczyk, K. L. J. Rygl, S. Sánchez, D. Sánchez-Arguelles, M. Sasada, T. Savolainen, F. P. Schloerb, K.-F. Schuster, L. Shao, Z. Shen, D. Small, B. W. Sohn, J. SooHoo, H. Sun, F. Tazaki, A. J. Tetarenko, P. Tiede, R. P. J. Tilanus, M. Titus, K. Toma, P. Torne, T. Trent, E. Traianou, S. Trippe, I. van Bemmel, H. J. van Langevelde, D. R. van Rossum, J. Wagner, D. Ward-Thompson, J. Wardle, J. Weintroub, N. Wex, R. Wharton, M. Wielgus, G. N. Wong, Q. Wu, D. Yoon, A. Young, K. Young, Z. Younsi, F. Yuan, Y.-F. Yuan, J. A. Zensus, G.-Y. Zhao, S.-S. Zhao, First M87 Event Horizon Telescope Results. VIII. Magnetic Field Structure near The Event Horizon. *Astrophys. J.* **910**, L13 (2021).

13. F. Yuan, H. Wang, H. Yang, The accretion flow in M87 is really MAD. *Astrophys. J.* **924**, 124 (2022).

14. M. Mościbrodzka, H. Falcke, H. Shiokawa, General relativistic magnetohydrodynamical simulations of the jet in M 87. *Astron. Astrophys.* **586**, A38 (2016).

15. I. V. Igumenshchev, Magnetically arrested disks and the origin of poynting jets: A numerical study. *Astrphys. J.* **677**, 317–326 (2008).

16. J. Dexter, A. Tchekhovskoy, A. Jiménez-Rosales, S. M. Ressler, M. Bauböck, Y. Dallilar, P. T. de Zeeuw, F. Eisenhauer, S. von Fellenberg, F. Gao, R. Genzel, S. Gillessen, M. Habibi, T. Ott, J. Stadler, O. Straub, F. Widmann, Sgr A* near-infrared flares from reconnection events in a magnetically arrested black hole accretion discs. *Mon. Not. R. Astron. Soc.* **497**, 4999–5007 (2020).

17. O. Porth, Y. Mizuno, Z. Younsi, C. M. Fromm, Flares in the Galactic Centre - I. Orbiting flux tubes in magnetically arrested black hole accretion discs. *Mon. Not. R. Astron. Soc.* **502**, 2023–2032 (2021).

18. B. Ripperda, M. Liska, K. Chatterjee, G. Musoke, A. A. Philippov, S. B. Markoff, A. Tchekhovskoy, Z. Younsi, Black hole flares: Ejection of accreted magnetic flux through 3D plasmoid-mediated reconnection. *Astrophys. J.* **924**, L32 (2022).

19. K. Chatterjee, R. Narayan, Flux eruption events drive angular momentum transport in magnetically arrested accretion flows. *Astrophys. J.* **941**, 30 (2022).

20. F. Guo, H. Li, W. Daughton, Y.-H. Liu, Formation of hard power laws in the energetic particle spectra resulting from relativistic magnetic reconnection. *Phys. Rev. Lett.* **113**, 155005 (2014).

21. L. Sironi, A. Spitkovsky, Relativistic reconnection: An efficient source of non-thermal particles. *Astrophys. J.* **783**, L21 (2014).

22. D. Ball, L. Sironi, F. Özel, Electron and proton acceleration in trans-relativistic magnetic reconnection: Dependence on plasma beta and magnetization. *Astrophys. J.* **862**, 80 (2018).

23. G. R. Werner, D. A. Uzdensky, M. C. Begelman, B. Cerutti, K. Nalewajko, Non-thermal particle acceleration in collisionless relativistic electron-proton reconnection. *Mon. Not. R. Astron. Soc.* **473**, 4840–4861 (2018).

24. X. Li, F. Guo, Y.-H. Liu, H. Li, A model for nonthermal particle acceleration in relativistic magnetic reconnection. *Astrophys. J.* **954**, L37 (2023).

25. E. Petersen, C. Gammie, Non-thermal models for infrared flares from Sgr A*. *Mon. Not. R. Astron. Soc.* **494**, 5923–5935 (2020).

26. A. Cruz-Osorio, C. Fromm, Y. Mizuno, A. Nathanail, Z. Younsi, O. Porth, J. Davelaar, H. Falcke, M. Kramer, L. Rezzolla, State-of-the-art energetic and morphological modelling of the launching site of the M87 jet. *Nat. Astron.* **6**, 103–108 (2022).

27. M. Mościbrodzka, C. F. Gammie, IPOLE - semi-analytic scheme for relativistic polarized radiative transport. *Mon. Not. R. Astron. Soc.* **475**, 43–54 (2018).

28. A. Chael, R. Narayan, M. D. Johnson, Two-temperature, magnetically arrested disc simulations of the jet from the supermassive black hole in M87. *Mon. Not. R. Astron. Soc.* **486**, 2873–2895 (2019).

29. C. J. White, J. M. Stone, C. F. Gammie, An Extension of the Athena++ Code Framework for GRMHD Based on Advanced Riemann Solvers and Staggered-mesh Constrained Transport. *Astrophys. J., Suppl. Ser.* **225**, 22 (2016).

30. J. M. Stone, K. Tomida, C. J. White, K. G. Felker, The Athena++ Adaptive Mesh Refinement Framework: Design and Magnetohydrodynamic Solvers. *Astrophys. J., Suppl. Ser.* **249**, 4 (2020).

31. B. Einfeldt, On Godunov-Type Methods for Gas Dynamics. *SIAM J. Numer. Anal.* **25**, 294–318 (1988).

32. O. Porth, K. Chatterjee, R. Narayan, C. F. Gammie, Y. Mizuno, P. Anninos, J. G. Baker, M. Bugli, C. Chan, J. Davelaar, L. Del Zanna, Z. B. Etienne, P. C. Fragile, B. J. Kelly, M. Liska, S. Markoff, J. C. McKinney, B. Mishra, S. C. Noble, H. Olivares, B. Prather, L. Rezzolla, B. R. Ryan, J. M. Stone, N. Tomei, C. J. White, Z. Younsi, K. Akiyama, A. Alberdi, W. Alef, K. Asada, R. Azulay, A.-K. Baczko, D. Ball, M. Baloković, J. Barrett, D. Bintley, L. Blackburn, W. Boland, K. L. Bouman, G. C. Bower, M. Bremer, C. D. Brinkerink, R. Brissenden, S. Britzen, A. E. Broderick, D. Broguiere, T. Bronzwaer, D.-Y. Byun, J. E. Carlstrom, A. Chael, S. Chatterjee, M.-T. Chen, Y. Chen, I. Cho, P. Christian, J. E. Conway, J. M. Cordes, G. B. Crew, Y. Cui, M. De Laurentis, R. Deane, J. Dempsey, G. Desvignes, S. S. Doeleman, R. P. Eatough,

H. Falcke, V. L. Fish, E. Fomalont, R. Fraga-Encinas, B. Freeman, P. Friberg, C. M. Fromm, J. L. Gómez, P. Galison, R. García, O. Gentaz, B. Georgiev, C. Goddi, R. Gold, M. Gu, M. Gurwell, K. Hada, M. H. Hecht, R. Hesper, L. C. Ho, P. Ho, M. Honma, C.-W. Huang, L. Huang, D. H. Hughes, S. Ikeda, M. Inoue, S. Issaoun, D. J. James, B. T. Jannuzi, M. Janssen, B. Jeter, W. Jiang, M. D. Johnson, S. Jorstad, T. Jung, M. Karami, R. Karuppusamy, T. Kawashima, G. K. Keating, M. Kettenis, J.-Y. Kim, J. Kim, J. Kim, M. Kino, J. Y. Koay, P. M. Koch, S. Koyama, M. Kramer, C. Kramer, T. P. Krichbaum, C.-Y. Kuo, T. R. Lauer, S.-S. Lee, Y.-R. Li, Z. Li, M. Lindqvist, K. Liu, E. Liuzzo, W.-P. Lo, A. P. Lobanov, L. Loinard, C. Lonsdale, R.-S. Lu, N. R. MacDonald, J. Mao, D. P. Marrone, A. P. Marscher, I. Martí-Vidal, S. Matsushita, L. D. Matthews, L. Medeiros, K. M. Menten, I. Mizuno, J. M. Moran, K. Moriyama, M. Mościbrodzka, C. Müller, H. Nagai, N. M. Nagar, M. Nakamura, R. Narayan, I. Natarajan, R. Neri, C. Ni, A. Noutsos, H. Okino, T. Oyama, F. Özel, D. C. M. Palumbo, N. Patel, U.-L. Pen, D. W. Pesce, V. Piétu, R. Plambeck, A. PopStefanija, J. A. Preciado-López, D. Psaltis, H.-Y. Pu, V. Ramakrishnan, R. Rao, M. G. Rawlings, A. W. Raymond, B. Ripperda, F. Roelofs, A. Rogers, E. Ros, M. Rose, A. Roshanineshat, H. Rottmann, A. L. Roy, C. Ruszczyk, K. L. J. Rygl, S. Sánchez, D. Sánchez-Arguelles, M. Sasada, T. Savolainen, F. P. Schloerb, K.-F. Schuster, L. Shao, Z. Shen, D. Small, B. W. Sohn, J. SooHoo, F. Tazaki, P. Tiede, R. P. J. Tilanus, M. Titus, K. Toma, P. Torne, T. Trent, S. Trippe, S. Tsuda, I. van Bemmel, H. J. van Langevelde, D. R. van Rossum, J. Wagner, J. Wardle, J. Weintroub, N. Wex, R. Wharton, M. Wielgus, G. N. Wong, Q. Wu, K. Young, Z. Younsi, F. Yuan, Y.-F. Yuan, J. A. Zensus, G. Zhao, S.-S. Zhao, Z. Zhu; Event Horizon Telescope Collaboration, The event horizon general relativistic magnetohydrodynamic code comparison project. *Astrophys. J. Supp.* **243**, 26 (2019).

33. J. C. McKinney, C. F. Gammie, A Measurement of the electromagnetic luminosity of a Kerr black Hole. *Astrophys. J.* **611**, 977–995 (2004).

34. R. Narayan, I. V. Igumenshchev, M. A. Abramowicz, Magnetically arrested disk: An energetically efficient accretion flow. *Publ. Astron. Soc. Jpn.* **55**, L69–L72 (2003).

35. R. Narayan, A. Sądowski, R. F. Penna, A. K. Kulkarni, GRMHD simulations of magnetized advection-dominated accretion on a non-spinning black hole: Role of outflows. *Mon. Not. R. Astron. Soc.* **426**, 3241–3259 (2012).

36. L. G. Fishbone, V. Moncrief, Relativistic fluid disks in orbit around Kerr black holes. *Astrophys. J.* **207**, 962–976 (1976).

37. R. F. Penna, A. Kulkarni, R. Narayan, A new equilibrium torus solution and GRMHD initial conditions. *Astron. Astrophys.* **559**, A116 (2013).

38. C. J. White, F. Chrystal, The effects of resolution on black hole accretion simulations of jets. *Astrophys. J.* **498**, 2428–2439 (2020).

39. J. F. Hawley, X. Guan, J. H. Krolik, Assessing quantitative results in accretion simulations: From local to global. *Astrophys. J.* **738**, 84 (2011).

40. R.-S. Lu, K. Asada, T. P. Krichbaum, J. Park, F. Tazaki, H.-Y. Pu, M. Nakamura, A. Lobanov, K. Hada, K. Akiyama, J.-Y. Kim, I. Martí-Vidal, J. L. Gomez, T. Kawashima, F. Yuan, E. Rose, W. Alef, S. Britzen, M. Bremer, P. T. P. Ho, M. Honma, D. H. Hughes, M. Inoue, W. Jiang, M. Kino, S. Koyama, M. Lindqvist, J. Liu, A. P. Marscher, S. Matsushita, H. Nagai, H. Rottmann, T. Savolainen, K.-F. Schuster, Z.-Q. Shen, P. de Vicente, R. C. Walker, H. Yang, J. A. Zensus, J. C. Algaba, A. Allardi, U. Bach, R. Berthold, D. Bintley, D.-Y. Byun, C. Casadio, S.-H. Chang, C.-C. Chang, S.-C. Chang, C.-C. Chen, M.-T. Chen, R. Chilson, T. C. Chuter, J. Conway, G. B. Crew, J. T. Dempsey, S. Dornbusch, A. Faber, P. Friberg, J. G. Garcia, M. G. Garrido, C.-C. Han, K.-C. Han, Y. Hasegawa, R. Herrero-Illana, Y.-D. Huang, C.-W. L. Huang, V. Impellizzeri, H. Jiang, H. Jinchi, T. Jung, J. Kallunki, P. Kirves, K. Kimura, J. Y. Koay, P. M. Koch, C. Kramer, A. Kraus, D. Kubo, C.-Y. Kuo, C.-T. Li, L.C.-C. Lin, C.-T. Liu, K.-Y. Liu, W.-P. Lo, L.-M. Lu, N. MacDonald, P. Martin-Cocher, H. Messias, Z. Meyer-Zhao, A. Minter, D. G. Nair, H. Nishioka, T. J. Norton, G. Nystrom, H. Ogawa, P. Oshiro, N. A. Patel, U.-L. Pen, Y. Pidopryhora, N. Pradel, P. A. Raffin, R. Rao, I. Ruiz, S. Sanchez, P. Shaw, W. Snow, T. K. Sridharan, R. Srinivasan, B. Tercero, P. Torne, E. Traianou, J. Wagner, C. Walther, T.-S. Wei, J. Yang, C.-Y. Yu, A ring-like accretion structure in M87 connecting its black hole and jet. *Nature.* **616**, 686–690 (2023).

41. Q. Zhang, F. Guo, W. Daughton, H. Li, X. Li, Efficient nonthermal ion and electron acceleration enabled by the flux-rope kink instability in 3D nonrelativistic magnetic reconnection. *Phys. Rev. Lett.* **127**, 185101 (2021).

42. F. Özel, D. Psaltis, R. Narayan, Hybrid thermal-nonthermal synchrotron emission from hot accretion flows. *Astrophys. J.* **541**, 234–249 (2000).

43. F. Yuan, E. Quataert, R. Narayan, Nonthermal electrons in radiatively inefficient accretion flow models of Sagittarius A*. *Astrophys. J.* **598**, 301–312 (2003).

44. J. Dexter, J. C. McKinney, E. Agol, The size of the jet launching region in M87. *Mon. Not. R. Astron. Soc.* **421**, 1517–1528 (2012).

45. K. Chatterjee, S. Markoff, J. Neilsen, Z. Younsi, G. Witzel, A. Tchekhovskoy, D. Yoon, A. Ingram, M. van der Klis, H. Boyce, T. Do, D. Haggard, M. A. Nowak, General relativistic MHD simulations of non-thermal flaring in Sagittarius A*. *Mon. Not. R. Astron. Soc.* **507**, 5281–5302 (2021).

46. A. E. Broderick, V. L. Fish, M. D. Johnson, K. Rosenfeld, C. Wang, S. S. Doeleman, K. Akiyama, T. Johannsen, A. L. Roy, Modeling seven years of event horizon telescope observations with radiatively inefficient accretion flow models. *Astrophys. J.* **820**, 137 (2016).

47. J. Davelaar, M. Mościbrodzka, T. Bronzwaer, H. Falcke, General relativistic magnetohydrodynamicalc-jet models for Sagittarius A*. *Astron. Astrophys.* **612**, A34 (2018).

48. A. E. Broderick, J. C. McKinney, Parsec-scale faraday rotation measures from general relativistic magnetohydrodynamic simulations of active galactic nucleus jets. *Astrophys. J.* **725**, 773 (2010).

49. J. Davelaar, H. Olivares, O. Porth, T. Bronzwaer, M. Janssen, T. Roelofs, Y. Mizuno, C. M. Fromm, H. Falcke, L. Rezzolla, Modeling non-thermal emission from the jet-launching region of M 87 with adaptive mesh refinement. *Astron. Astrophys.* **632**, A2 (2019).

50. D. Ball, F. Özel, D. Psaltis, C. K. Chan, L. Sironi, The properties of reconnection current sheets in GRMHD simulations of radiatively inefficient accretion flows. *Astrophys. J.* **853**, 184 (2018).

51. C. B. Singh, Y. Mizuno, E. M. de Gouveia Dal Pino, Spatial growth of current-driven instability in relativistic rotating jets and the search for magnetic reconnection. *Astrophys. J.* **824**, 48 (2016).

52. E. P. Alves, J. Zrake, F. Fiuza, Efficient nonthermal particle acceleration by the kink instability in relativistic jets. *Phys. Rev. Lett.* **121**, 245101 (2018).

53. T. E. Medina-Torrejón, E. M. de Gouveia Dal Pino, L. H. S. Kadowaki, G. Kowal, C. B. Singh, Y. Mizuno, Particle acceleration by relativistic magnetic reconnection driven by kink instability turbulence in poynting flux-dominated jets. *Astrophys. J.* **908**, 193 (2021).

54. A. Tomimatsu, T. Matsuoka, M. Takahashi, Screw instability in black hole magnetospheres and a stabilizing effect of field-line rotation. *Phys.Rev. D* **64**, 123003 (2001).

55. M. Nakamura, H. Li, S. T. Li, Stability properties of magnetic tower jets. *Astrophys. J.* **656**, 721–732 (2007).

56. H. Yang, F. Yuan, Y. F. Yuan, C. White, Numerical simulation of hot accretion flows (IV): Effects of black hole spin and magnetic field strength on the wind and the comparison between wind and jet properties. *Astrophys. J.* **914**, 131 (2021).

57. K. Nishikawa, Y. Mizuno, J. L. Gomez, I. Dutan, J. Niemiec, O. Kobzar, N. MacDonald, A. Meli, M. Pohl, K. Hirotani, Rapid particle acceleration due to recollimation shocks and turbulent magnetic fields in injected jets with helical magnetic fields. *Mon. Not. R. Astron. Soc.* **493**, 2652–2658 (2020).

58. L. Sironi, M. E. Rowan, R. Narayan, Reconnection-driven particle acceleration in relativistic shear flows. *Astrophys. J.* **907**, L44 (2021).

59. B. R. Ryan, S. M. Ressler, J. C. Dolence, C. Gammie, E. Quataert, Two-temperature GRRMHD simulations of M87. *Astrophys. J.* **864**, 126 (2018).

60. Event Horizon Telescope Collaboration, K. Akiyama, A. Alberdi, W. Alef, K. Asada, R. Azulay, A.-K. Baczko, D. Ball, M. Baloković, J. Barrett, D. Bintley, L. Blackburn, W. Boland, K. Bouman, G. C. Bower, M. Bremer, C. D. Brinkerink, R. Brissenden, S. Britzen, A. E. Broderick, D. Broguiere, T. Bronzwaer, D.-Y. Byun, J. E. Carlstrom, A. Chael, C. Chan, S. Chatterjee, K. Chatterjee, M.-T. Chen, Y. Chen, I. Cho, P. Christian, J. E. Conway, J. M. Cordes, G. B. Crew, Y. Cui, J. Davelaar, M. De Laurentis, R. Deane, J. Dempsey, G. Desvignes, J. Dexter, S. S. Doeleman, R. P. Eatough, H. Falcke, V. L. Fish, E. Fomalont, R. Fraga-Encinas, P. Friberg, C. M. Fromm, J. L. Gómez, P. Galison, C. F. Gammie, R. García, O. Gentaz, B. Georgiev, C. Goddi, R. Gold, M. Gu, M. Gurwell, K. Hada, M. H. Hecht, R. Hesper, L. C. Ho, P. Ho, M. Honma, C.-W. L. Huang, L. Huang, D. H. Hughes, S. Ikeda, M. Inoue, S. Issaoun, D. J. James, B. T. Jannuzi, M. Janssen, B. Jeter, W. Jiang, M. D. Johnson, S. Jorstad, T. Jung, M. Karami, R. Karuppusamy, T. Kawashima, G. K. Keating, M. Kettenis, J.-Y. Kim, J. Kim, J. Kim, M. Kino, J. Y. Koay, P. M. Koch, S. Koyama, M. Kramer, C. Kramer, T. P. Krichbaum, C.-Y. Kuo, T. R. Lauer, S.-S. Lee, Y.-R. Li, Z. Li, M. Lindqvist, K. Liu, E. Liuzzo, W.-P. Lo, A. P. Lobanov, L. Loinard, C. Lonsdale, R.-S. Lu, N. R. MacDonald, J. Mao, S. Markoff, D. P. Marrone, A. P. Marscher, I. Martí-Vidal, S. Matsushita, L. D. Matthews, L. Medeiros, K. M. Menten, Y. Mizuno, I. Mizuno, J. M. Moran, K. Moriyama, M. Moscibrodzka, C. Müller, H. Nagai, N. M. Nagar, M. Nakamura, R. Narayan, G. Narayanan, I. Natarajan, R. Neri, C. Ni, A. Noutsos, H. Okino, H. Olivares, T. Oyama, F. Özel, D. C. M. Palumbo, N. Patel, U.-L. Pen, D. W. Pesce, V. Piétu, R. Plambeck, A. PopStefanija, O. Porth, B. Prather, J. A. Preciado-López, D. Psaltis, H.-Y. Pu, V. Ramakrishnan, R. Rao, M. G. Rawlings, A. W. Raymond, L. Rezzolla, B. Ripperda, F. Roelofs, A. Rogers, E. Ros, M. Rose, A. Roshanineshat, H. Rottmann, A. L. Roy, C. Ruszczyk, B. R. Ryan, K. L. J. Rygl, S. Sánchez, D. Sánchez-Arguelles, M. Sasada, T. Savolainen, F. P. Schloerb, K.-F. Schuster, L. Shao, Z. Shen, D. Small, B. W. Sohn, J. SooHoo, F. Tazaki, P. Tiede, R. P. J. Tilanus, M. Titus, K. Toma, P. Torne, T. Trent, S. Trippe, S. Tsuda, I. van Bemmel, H. J. van Langevelde, D. R. van Rossum, J. Wagner, J. Wardle, J. Weintroub, N. Wex, R. Wharton, M. Wielgus, G. N. Wong, Q. Wu, A. Young, K. Young, Z. Younsi, F. Yuan, Y.-F. Yuan, J. A. Zensus, G. Zhao, S.-S. Zhao, Z. Zhu, J. Anczarski, F. K. Baganoff, A. Eckart, J. R. Farah, D. Haggard, Z. Meyer-Zhao, D. Michalik, A. Nadolski, J. Neilsen, H. Nishioka, M. A. Nowak, N. Pradel, R. A. Primiani, K. Souccar, L. Vertatschitsch, P. Yamaguchi, S. Zhang, First M87 event horizon telescope results. V. physical origin of the asymmetric ring. *Astrophys. J.* **875**, L5 (2019).

61. K. Gebhardt, J. Adams, D. Richstone, T. R. Lauer, S. M. Faber, K. Gultekin, J. Murphy, S. Tremaine, The black hole mass in M87 from Gemini/NIFS adaptive optics observations. *Astrophys. J.* **729**, 119 (2011).

62. R. C. Walker, P. E. Hardee, F. B. Davies, C. Ly, W. Junor, The structure and dynamics of the subparsec jet in M87 based on 50 VLBA observations over 17 years at 43 GHz. *Astrophys. J.* **855**, 128 (2018).

63. K. Asada, M. Nakamura, The structure of the M87 jet: A transition from parabolic to conical streamlines. *Astrophys. J.* **745**, L28 (2012).

64. J.-Y. Kim, T. P. Krichbaum, R.-S. Lu, E. Ros, U. Bach, M. Bremer, P. de Vicente, M. Lindqvist, J. A. Zensus, The limb-brightened jet of M87 down to the 7 Schwarzschild radii scale. *Astron. Astrophys.* **616**, A188 (2018).

65. M. Janssen, C. Goddi, I. M. van Bemmel, M. Kettenis, D. Small, E. Liuzzo, K. Rygl, I. Martí-Vidal, L. Blackburn, M. Wielgus, H. Falcke, rPICARD: A CASA-based calibration pipeline for VLBI data. Calibration and imaging of 7 mm VLBA observations of the AGN jet in M 87. arXiv: 1902.01749 [astro-ph.IM] (15 May 2019).

66. M. Nakamura, K. Asada, K. Hada, H.-Y. Pu, S. Noble, C. Tseng, K. Toma, M. Kino, H. Nagai, K. Takahashi, J.-C. Algaba, M. Orienti, K. Akiyama, A. Doi, G. Giovannini, M. Giroletti, M. Honma, S. Koyama, R. Lico, K. Niinuma, F. Tazaki, Parabolic Jets from the Spinning Black Hole in M87. *Astrophys. J.* **868**, 146 (2018).

67. K. I. Kellermann, M. L. Lister, D. C. Homan, R. C. Vermeulen, M. H. Cohen, E. Ros, M. Kadler, J. A. Zensus, Y. Y. Kovalev, Sub-milliarcsecond imaging of quasars and active galactic nuclei. III. kinematics of parsec-scale radio jets. *Astrophys. J.* **609**, 539–563 (2004).

68. Y. Y. Kovalev, M. L. Lister, D. C. Homan, K. I. Kellermann, The inner jet of the radio galaxy M87. *Astrophys. J.* **668**, L27–L30 (2007).

69. K. Hada, M. Kino, A. Doi, H. Nagai, M. Honma, K. Akiyama, F. Tazaki, R. Lico, M. Giroletti, G. Giovannini, M. Orienti, Y. Hagiwara, High-sensitivity 86 GHz (3.5 mm) VLBI observations of M87: Deep imaging of the jet base at a resolution of 10 Schwarzschild radii. *Astrophys. J.* **817**, 131 (2016).

70. K. Hada, J. H. Park, M. Kino, K. Niinuma, B. W. Sohn, H. W. Ro, T. Jung, J.-C. Algaba, G.-Y. Zhao, S.-S. Lee, K. Akiyama, S. Trippe, K. Wajima, S. Sawada-Satoh, F. Tazaki, I. Cho, J. Hodgson, J. A. Lee, Y. Hagiwara, M. Honma, S. Koyama, J. Oh, T. Lee, H. Yoo, N. Kawaguchi, D.-G. Roh, S.-J. Oh, J.-H. Yeom, D.-K. Jung, C. Oh, H.-R. Kim, J.-Y. Hwang, D.-Y. Byun, S.-H. Cho, H.-G. Kim, N. Kobayashi, K. M. Shibata, Pilot KaVA monitoring on the M 87 jet: Confirming the inner jet structure and superluminal motions at sub-pc scales. *Publ. Astron. Soc. Jpn.* **69**, 71 (2017).

71. F. Mertens, A. P. Lobanov, R. C. Walker, P. E. Hardee, Kinematics of the jet in M 87 on scales of 100-1000 Schwarzschild radii. *Astron. Astrophys.* **595**, A54 (2016).

72. J. Park, K. Hada, M. Kino, M. Nakamura, J. Hodgson, H. Ro, Y. Cui, K. Asada, J.-C. Algaba, S. Sawada-Satoh, S.-S. Lee, I. Cho, Z. Shen, W. Jiang, S. Trippe, K. Niinuma, B. W. Sohn, T. Jung, G.-Y. Zhao, K. Wajima, F. Tazaki, M. Honma, T. An, K. Akiyama, D.-Y. Byun, J. Kim, Y. Zhang, X. Cheng, H. Kobayashi, K. M. Shibata, J. W. Lee, D.-G. Roh, S.-J. Oh, J.-H. Yeom, D.-K. Jung, C. Oh, H.-R. Kim, J.-Y. Hwang, Y. Hagiwara, Kinematics of the M 87 Jet in the collimation zone: Gradual acceleration and velocity stratification. *Astrophys. J.* **887**, 147 (2019).

73. E. Kravchenko, M. Giroletti, K. Hada, D. L. Meier, M. Nakamura, J. Park, R. C. Walker, Linear polarization in the nucleus of M87 at 7 mm and 1.3 cm. *Astron. Astrophys.* **637**, L6 (2020).

74. J. Park, K. Hada, M. Kino, M. Nakamura, H. Ro, S. Trippe, Faraday rotation in the Jet of M87 inside the Bondi radius: Indication of winds from hot accretion flows confining the relativistic jet. *Astrophys. J.* **871**, 257 (2019).

75. V. S. Beskin, Y. N. Istomin, V. I. Parev, Filling the magnetosphere of a supermassive black hole with plasma. *Sov. Astron.* **36**, 642 (1992).

76. A. Levinson, F. Rieger, Variable TeV EMISSION as a manifestation of jet formation in M87? *Astrophys. J.* **730**, 123 (2011).

77. A. E. Broderick, A. Tchekhovskoy, Horizon-scale lepton acceleration in jets: Explaining the compact radio emission in M87. *Astrophys. J.* **809**, 97 (2015).

78. A. A. Zdziarski, D. G. Phuravhathu, M. Sikora, M. Böttcher, J. O. Chibueze, The composition and power of the jet of the broad-line radio galaxy 3C 120. *Astrophys. J.* **928**, LV (2022).

79. E. E. Nokhrina, V. S. Beskin, Y. Y. Kovalev, A. A. Zheltoukhov, Intrinsic physical conditions and structure of relativistic jets in active galactic nuclei. *Mon. Not. R. Astron. Soc.* **447**, 2726–2737 (2015).

**Acknowledgments:** We thank C. White for the help on numerical simulations and A. Tchekhovskoy for the discussions on pair cascades in black hole magnetosphere and jet physics. The anonymous referees are acknowledged for their very constructive suggestions and comments, which have substantially improved the paper. The computations in this work were run on the Pi-2.0 and Siyuan-1 cluster supported by the Center for High Performance Computing at Shanghai Jiao Tong University, the Supercomputer Cluster at Shanghai Supercomputer Center, and the High Performance Computing Resource in the Core Facility for Advanced Research Computing at Shanghai Astronomical Observatory, Chinese Academy of Sciences. **Funding:** F.Y. and H.Y. are supported by Natural Science Foundation of China (grants 12133008, 12192220, and 12192223) and China Manned Space Project (CMS-CSST-2021-B02). H.L. and F.G. are supported by the U.S. Department of Energy Fusion of Energy Science. Y.M. is supported by the National Natural Science Foundation of China (grant 12273022) and the Science and Technology Commission of Shanghai Municipality orientation program of basic research for international scientists (grant no. 22JC1410600). R.L. is supported by the Max Planck Partner Group of the MPG and the CAS and Natural Science Foundation of China (grants 12325302 and 11933007), the Key Research Program of Frontier Sciences, CAS (grant ZDBS-LY-SLH011), and the Shanghai Pilot Program for Basic Research–Chinese Academy of Science, Shanghai Branch (grant JCYJ-SHFY-2021-013). L.H. is supported by the Natural Science Foundation of China (grants 11721303, 11991052, and 12011540375) and the China Manned Space Project (CMS-CSST-2021-A04). A.A.Z. is supported by the Polish National Science Center under the grant 2019/35/B/ST9/03944. J.S.W. is supported by the Alexander von Humboldt Foundation. **Author contributions:** F.Y. designed and led the project and drafted the paper. Numerical simulations of black hole accretion, analysis to the simulation data, and calculations of radiative transfer were performed by H.Y. H.L., Y.M., F.G., and J.S.W. contributed to the analysis of the particle accelerations by magnetic reconnection in the jet. R.L. and L.H. contributed to the observational aspects of the M87 jet and their comparisons with theoretical predictions. The polarization calculation was performed by X.L. A.A.Z. contributed to the discussion and calculation of the pair cascades in the black hole magnetosphere. All authors were involved in the discussions and final version of the paper. **Competing interests:** The authors declare that they have no competing interests. **Data and materials availability:** All data needed to evaluate the conclusions in the paper are present in the paper and/or the Supplementary Materials. The ATHENA++ and IPOLE codes used in this work are publicly available at https://github.com/PrincetonUniversity/athena-public-version and https://github.com/AFD-Illinois/ipole. The simulation data and analysis code for the accretion flows and jet at 86GHz can be downloaded from https://cdsarc.u-strasbg.fr/viz-bin/qcat?J/A+A/616/A188.

Submitted 5 December 2023
Accepted 20 February 2024
Published 22 March 2024
10.1126/sciadv.adn3544

# Science Advances

**▲AAAS**

## Supplementary Materials for

### Modeling the inner part of the jet in M87: Confronting jet morphology with theory

Hai Yang *et al.*

Corresponding author: Feng Yuan, fyuan@fudan.edu.cn

**This PDF file includes:**

Supplementary Text

## S1 The effect of choosing different boundaries of the emission region in the jet

In our fiducial model, we set $\sigma_{cut} = 5$ to determine our boundary of the radiation region in the jet. In this section we examine the effects of choosing different boundaries.

We have first tried different values of $\sigma_{cut}$. Fig. S10 shows the predicted images at 86 GHz with $\sigma_{cut} = 3$ & 6, together with the image predicted by the fiducial $\sigma_{cut} = 5$ model. We can see from the figure that three images are very similar to each other. We have also compared the predicted jet width and found they are also similar. The similarity implies that our results are not very sensitive to the choice of $\sigma_{cut}$. Physically, the limb-brightening is because of the combination effects of the magnetic field distribution, the spatial distribution of nonthermal electrons, and the Lorentz factor of the jet. From the middle panel of Fig. 1, we can see that the region of large $N_{pl}$, where we may expect most of the radiation originates, deviates from the $\sigma_{cut} = 5$ line. This perhaps explains why the results are not very sensitive to the value of $\sigma_{cut}$. But we note that the results will be notably different if the value of $\sigma_{cut}$ deviates from $\sigma_{cut} = 5$ too much.

Usually, the motivation of choosing a value of $\sigma_{cut}$ to determine the radiation region is that we believe the thermodynamics outside of this region obtained in the simulation is no longer reliable. In this case, we should ensure that no matter escapes from this "unreliable'' region and enters the radiation region. To examine this issue, we have compared the streamline and the $\sigma_{cut} = 5$ line, as shown by Fig. S11. We can see from the figure that the $\sigma_{cut} = 5$ line is very close to the streamline, the two lines almost coincide with each other in most of the region.

Alternatively, instead of using $\sigma_{cut}$ to determine the boundary of the radiation region, we have used the streamline shown in Fig. S11 as the boundary of the radiation region in the jet. Fig. S12 shows the predicted image at 86 GHz in this case. By comparing this result with Fig. 2, we can see that they are very similar.

## S2 Additional observational constraints on the jet models: velocity field, power, and polarization of the M87 jet

Multiwavelength VLBI and optical observations provide both subluminal and superluminal features of the proper motion of the jet in M87. The observed velocity field, shown by the dots in Fig. S13, provides a constraint on the jet model (*66*). We examine our simulated velocity and compare with observations. The Lorentz factor in our simulation is calculated by $\Gamma = \sqrt{-g_{tt}}u^t$, which is measured in Boyer-Lindquist coordinates. From the relation between the Lorentz factor $\Gamma$ and velocity $\beta_v$, $\Gamma \equiv (1 - \beta_v^2)^{-1/2}$, we can calculate the predicted velocity $\beta_v$ of the jet plasma. The velocity obtained from observation is the apparent speed of the moving components in the jet, $\beta_{app}$. The intrinsic speed $\beta_v$ is calculated using the equation $\beta_v = \beta_{app}/(\beta_{app}cos\theta_v + sin\theta_v)$. The comparison of the value $\Gamma\beta_v$ between simulation and the observation is shown in Fig. S11. We can see that predicted velocity is consistent with observations.

The second constraint is the jet power. Due to the physical uncertainties in the models used to estimate the jet power and the wide range in length and timescales probed by the observations, a wide range of powers is obtained, from $10^{42}erg\ s^{-1}$ to $10^{45}erg\ s^{-1}$ (62). We calculate the predicted jet power following Ref. (*4*), $P_{BZ} = \frac{\kappa_0}{4\pi c}\Omega_H^2\Phi_{BH}^2f(\Omega_H)$, where $\kappa_0$ is numerical constant that depends on the geometry of the magnetic field and we adopt $\kappa_0$=0.044; $\Omega_H = ac/(2r_H)$; $\Phi_{BH}$ is the absolute magnetic flux threading one hemisphere of the black hole event horizon; $f(\Omega_H)$ is a modifying factor for high spin $a$, which is $f(\Omega_H) \approx 1 + 1.38(\Omega_H r_g/c)^2 - 9.2(\Omega_H r_g/c)^4$. The power of BZ-jet calculated in MAD98, MAD05, and SANE98 are $6.3\times10^{43}erg\ s^{-1}$, $3.1 \times 10^{43}erg\ s^{-1}$, $9.6\times10^{42}erg\ s^{-1}$, respectively. All are within the observed range.

The third constraint is polarization. High angular resolution polarimetric observations of the nucleus of M87 are performed using the Very Long Baseline Array at 24 GHz and 43 GHz, and polarization structures within ~0.7 (at 43 GHz) -1 (at 24 GHz) mas around the core region are obtained (*73*). The average linear polarization degree is about 2-3%. We have calculated the polarization degree predicted by our fiducial MAD98 model using the IPOLE code. In the calculation, we first get the Stokes I, Q, U and V at 43GHz, then convolve each component with the beam of $0.38 \times 0.17\ mas$ at $-10°$. The calculation result for the central nuclear region of M87 is shown in Fig. S14.

Comparing this result with that shown in Ref. (*73*), we can see that the predicted polarization degree is notably higher than the observed value. The reason is because our calculation only takes into account the depolarization due to the plasma within our simulation domain, i.e., $r \lesssim 1000 r_g$ from the black hole. The RM due to the plasma in this region is small because of the small inclination angle and the low gas density close to the jet axis. From the detailed theoretical modeling presented in Ref. (*13*), the RM contributed by the plasma outside the simulation domain, mainly the wind launched from the hot accretion flow in M87, can be as large as $\geq 10^5\ rad/m^2$; thus strong depolarization is expected there. This prediction is consistent with the result obtained by the VLBA observations (*74*), which show that the RM is of external origin and larger than $10^4\ rad/m^2$ at a de-projected distance of as large as $10^4 r_g$. The detailed calculation of the depolarization by this part of plasma is beyond the scope of the present paper. The high polarization degree we obtain therefore implies that our theoretical prediction is not in conflict with observations.

## S3 Images of jet predicted by other accretion modes and black hole spin

To show the effects of accretion mode and black hole spin, we have calculated the images corresponding to other three models, i.e., MAD05, MAD00, and SANE98. We have searched 30 (for MAD05), 20 (for SANE98), and 20 (for MAD00) snapshots from their simulation data after the models have reached their steady states and calculated their images. The representative results are shown in Fig. S15. For MAD05 and SANE98, the predicted jet extends along the jet axis up to 2 mas, similar to MAD98. However, both models fail to reproduce the limb-brightening feature. Moreover, the predicted jet width by the SANE98 model is too small compared to observations, as shown by Fig. 4.

To understand the reasons, we show in Fig. S16 the two-dimensional spatial distributions of $N_{pl}/N_{tot}$. For all three models, we find that, at distances not too far away from the black hole, say $z < 400r_g$, the large value of $N_{pl}/N_{tot}$ (thus high number density of accelerated nonthermal electrons) are distributed in a rather broad layer, not as narrow as in the case of MAD98. This may be the reason for the absence of limb-brightening in their predicted images. For SANE98, different from MAD98 and MAD05, the value of $N_{pl}/N_{tot}$ is also very high at the region very close to the jet axis. This explains why the jet width predicted by SANE98 is too small, as shown in Fig. 4. For MAD00, we find from the figure that the value of $N_{pl}/N_{tot}$ close to the black hole is notably smaller than the other three models. This implies there are fewer nonthermal electrons in the jet and may explain why almost no elongated jet-like structure is seen in the middle panel of Fig. S15. The results of the spatial distribution of $N_{pl}/N_{tot}$ in all four models, including their differences among them, should be caused by their specific configurations of the magnetic field lines, which in turn must be due to the differences of accretion mode and black hole spin.

**S4 Is it possible to put constraints on the pair cascade model?**

In addition to the electron-proton model presented in the present paper, there are alternative electron-positron models in the literature for the emission of the jet in M87 (*75-79*). In these models, pairs are produced by photon-photon collision processes and are accelerated by unscreened electric fields in the magnetosphere of the rotating black hole. The accelerated particles Compton up-scatter background photons, which collide with other background photons and produce a pair cascade. The radiation of these pairs could be responsible for the observed radiation of the jet. Such a process cannot be evaluated using ideal GRMHD simulation models as presented in this work. But whether our model can present some constraints on this mechanism?

There exists a critical charge number density in the magnetosphere below which a strong electric field forms and electrons can be accelerated and consequently a cascade of pair production occurs, i.e., the Goldreich-Julian density, $n_{GJ}$. Detailed calculations have shown that the number density of pairs produced by the pair cascade mechanism in the spark gap is $\sim 10^3 n_{GJ}$ (*76*). In the following, we estimate the value of $n_{GJ}$ and compare it with the number density obtained in our model.

Instead of directly comparing the number density, we compare the number flux. The underlying assumption is that, the pairs produced in the magnetosphere should be advected into the jet and the number flux should be conserved. The number flux of electrons corresponding to the Goldreich-Julian density is roughly estimated to be:

$$F_{GJ} \sim n_{GJ} r_g^2 v_r, \qquad\qquad (13)$$

where $n_{GJ} = \frac{\Omega B}{4\pi e c}$, $\Omega = \frac{ac^3}{4GM_{BH}}$, $r_g$ is assumed to be the radius of the gap, $v_r$ is the typical radial velocity of the pairs in the gap, and $e$ is the electron charge. On the other hand, in our model the number flux of electrons at a large radii $r$ within the BZ-jet is estimated to be:

$$F_e \approx 2\pi \int_{\theta_{BZ}} v_r n_e \Gamma r^2 sin(\theta) d\theta, \qquad (14)$$

where $n_e = n_i = \rho/m_p$, $\Gamma$ is the bulk motion Lorentz factor of the jet, $v_r \sim c$ is the typical radial velocity at $r$ in the jet.

We have calculated the above two integrals using the values obtained from $\phi$-averaged values of our fiducial MAD98 model. We find that $n_{GJ} = 4.1 \times 10^{-7} cm^{-3}$, $F_{GJ} \approx 10^{33} s^{-1}$. At $r \sim 500 \, r_g$, $F_e \approx 10^{45} s^{-1}$. So we have $\frac{F_e}{F_{GJ}} \sim 10^{12}$. This value roughly agrees with Ref. (79). Such a large ratio suggests that pair cascades in the black hole magnetosphere are likely not able to provide enough electrons to produce the observed radio flux of the jet in M87.

In addition to the number flux, another constraint comes from timescale. The radiative timescale of pairs produced in the black hole magnetosphere should be long enough so that they can propagate into a large distance of the jet and produce the radio emission there. The dynamical time is roughly $t_{dyn} = 1.6 \times 10^7 (\frac{z}{500 \, r_g}) \, s$, while the radiative timescale at the magnetosphere is roughly $t_{cool} = \frac{9 \, m_e{}^3 c^5}{4 e^4 B^2 \gamma_{min}} = 1.4 \times 10^3 (\frac{B}{74 G})^{-2} (\frac{\gamma_{min}}{100})^{-1} \, s$. The radiative timescale is much shorter than the dynamical timescale, which is another evidence against the pair produced in the black hole magnetosphere to be responsible for the observed radiation in the jet.

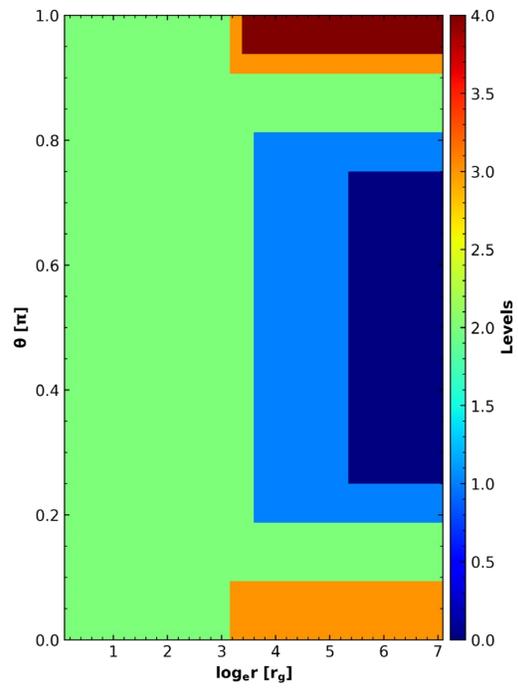

**Fig. S1. The final grid level for MAD98 in the** $log(r) - \theta/\pi$ **plane.**

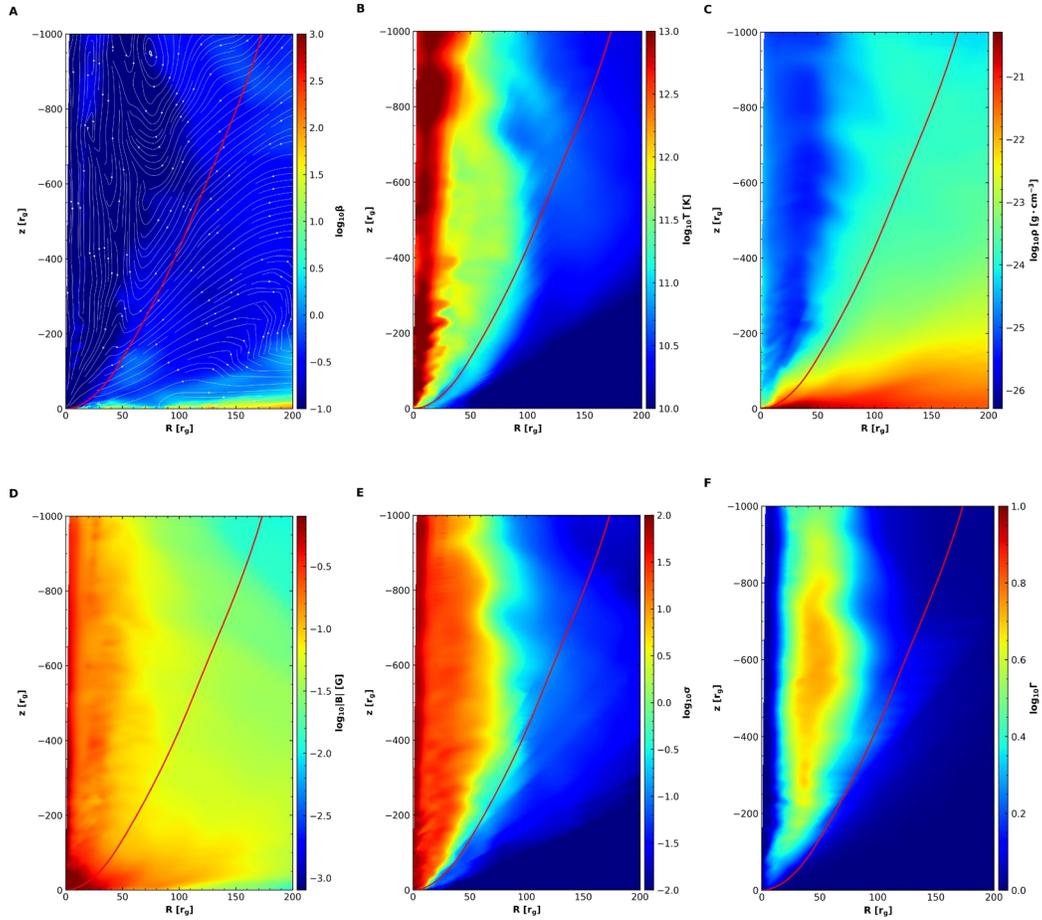

**Fig. S2. The $\phi$-averaged two-dimensional distribution of different physical quantities in R-z plane.** They are the (**A**) plasma $\beta$, (**B**) temperature, (**C**) number density, (**D**) magnetic field strength, (**E**) magnetization parameter, and (**F**) Lorentz factor of MAD98 at a simulation time of t=27,400, the same time as Fig. 1. The white curves in the top-left panel denote magnetic field lines. The red lines in all panels denote the $\phi$-averaged boundary of the BZ-jet.

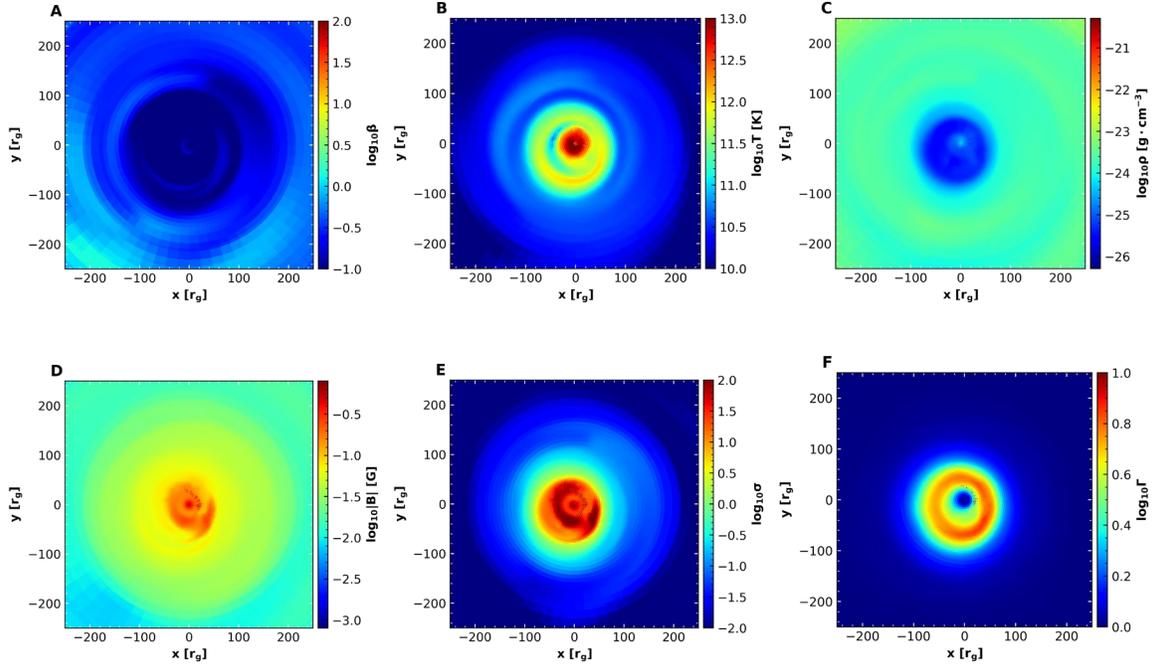

**Fig. S3. The two-dimensional distribution in the $x - y$ plane at $z = 600r_g$ of different physical quantities.** (**A**) plasma $\beta$, (**B**) temperature, (**C**) density, (**D**) magnetic field strength, (**E**) magnetization parameter, and (**F**) Lorentz factor of MAD98 at a simulation time of t=27,400, the same time as Fig. 1.

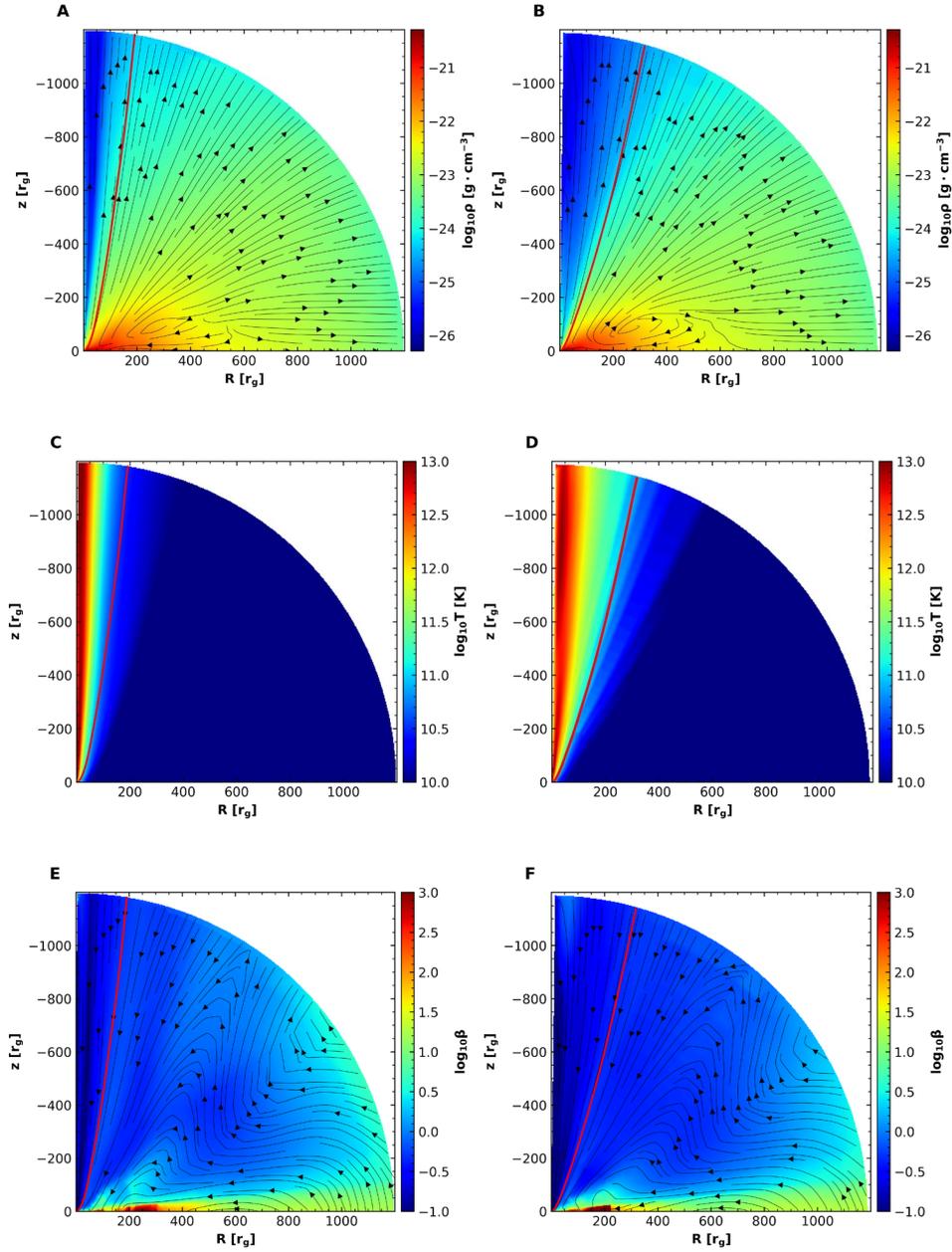

**Fig. S4. Comparison of time and φ-averaged physical quantities at high and low resolutions.**
(**A** and **B**) density, (**C** and **D**) temperature, and (**E** and **F**) plasma $\beta$ for high (left) and low (right) resolution simulations of MAD98. The red line in all panels denotes the φ-averaged boundary of the BZ-jet. The black curves with arrows in the top panel denote velocity field lines, the black curves with arrows in the bottom panel denote magnetic field lines.

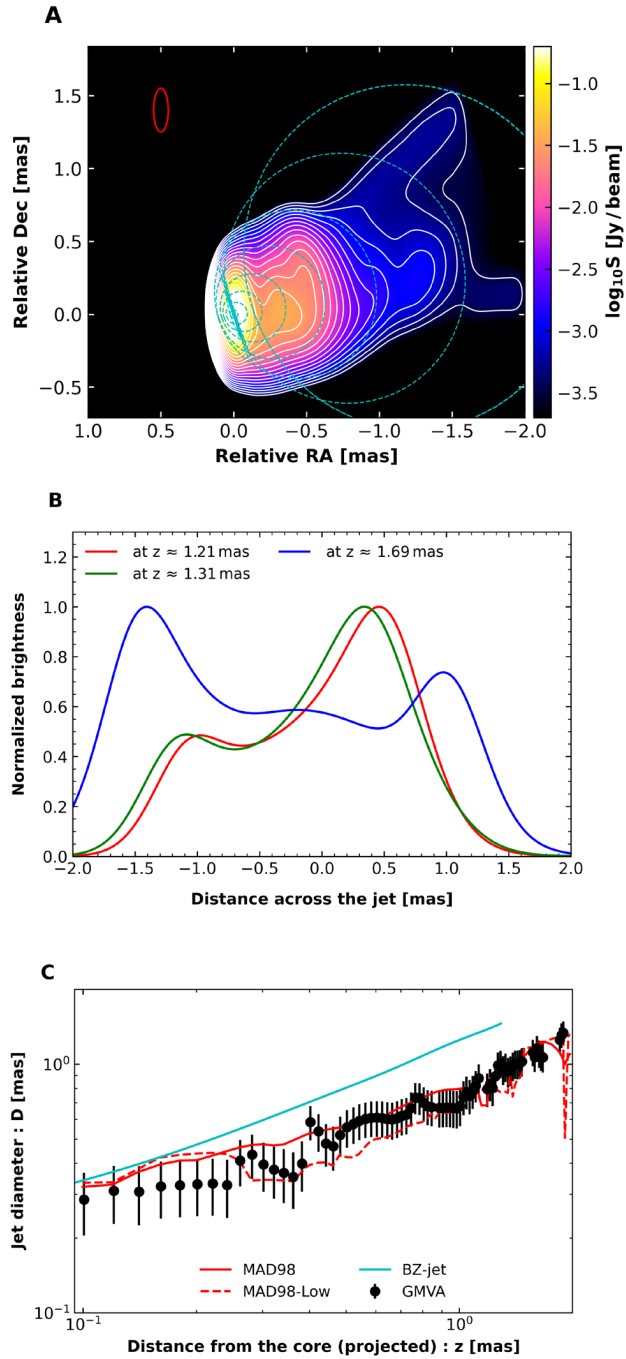

**Fig. S5. The predictions of the low-resolution MAD98 model.** (**A**) The images at 86 GHz. (**B**) limb-brightening feature. (**C**) the jet width. The dotted circles in the top panel denote the boundary of the BZ-jet.

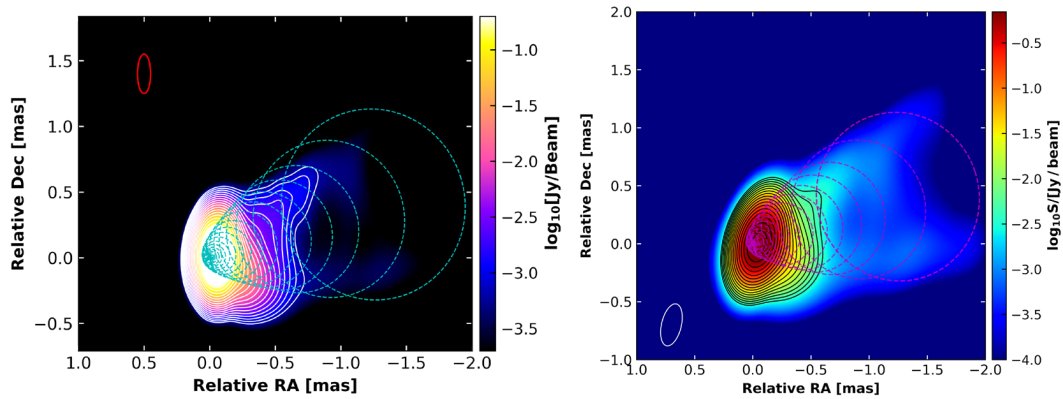

**Fig. S6.  Images predicted by a $N_{pl}/N_{tot} = 0.5$ test model.** (**A**) 86 GHz and (**B**) 43 GHz. All other model parameters are the same with our fiducial "current density" model. Comparison with Fig. 2 indicates that this model is more similar to the "thermal-only'' model rather than the "current density'' model.

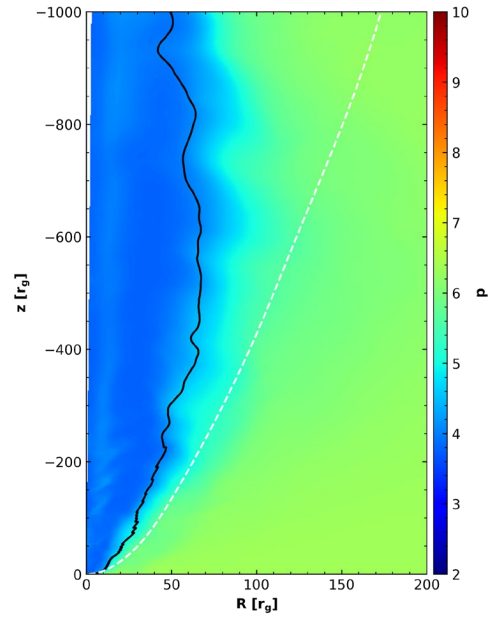

**Fig. S7. The distribution of the value of power-law index $p$ of electrons accelerated by magnetic reconnection**. It is calculated by equation (5) based on the simulation data of MAD98.

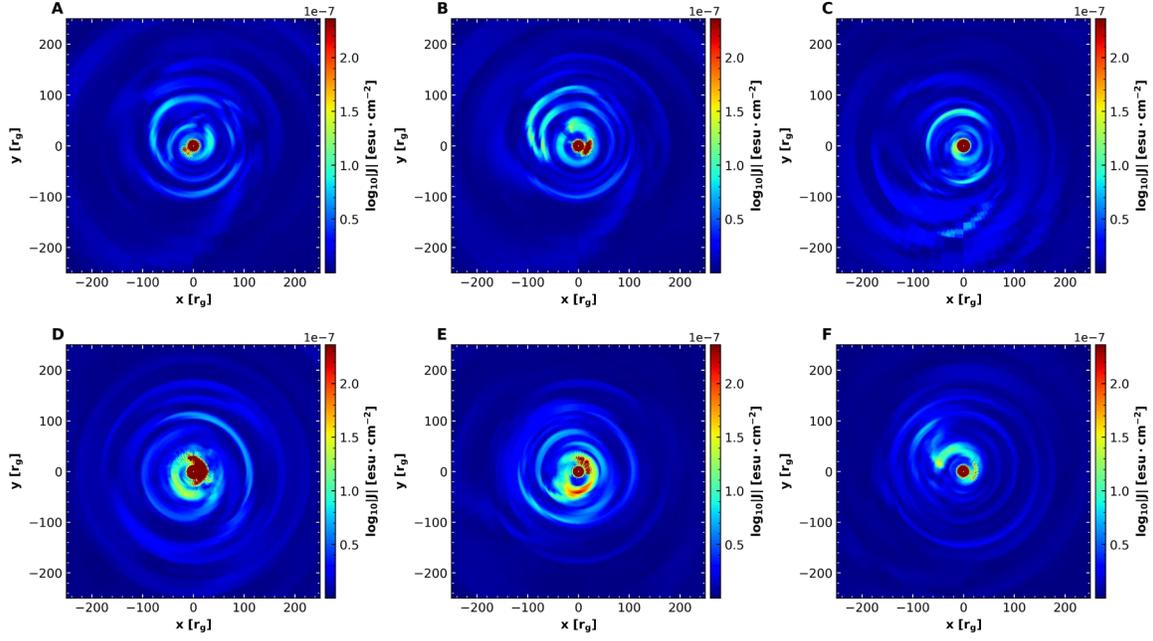

**Fig. S8. The distribution of the magnitude of current density in MAD98 at different simulation times in the $x - y$ plane at $z = 600 r_g$.** The corresponding times of panels (**A**) - (**F**) are t=25300, 25500, 26700, 27400, 33800, and 37300 $r_g/c$, respectively.

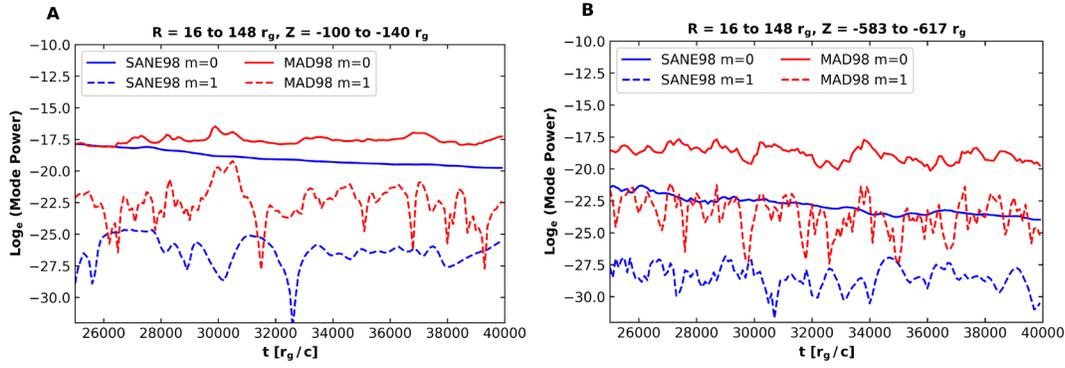

**Fig. S9. The time evolution of the azimuthal Fourier power in $m = 0$ and $m = 1$ modes for SANE98 (blue) and MAD98 (red).** From left to right correspond to two different regions of the jet. (**A**) At small $z \sim 120 r_g$. (**B**) At large $z \sim 600 r_g$. The larger power in MAD98 compared to SANE98 and the presence of notable $m = 1$ mode power even at small $z$ ($\sim 120 r_g$), among others, indicate that magnetic eruption rather than kink instability is likely the physical mechanism of driving reconnection.

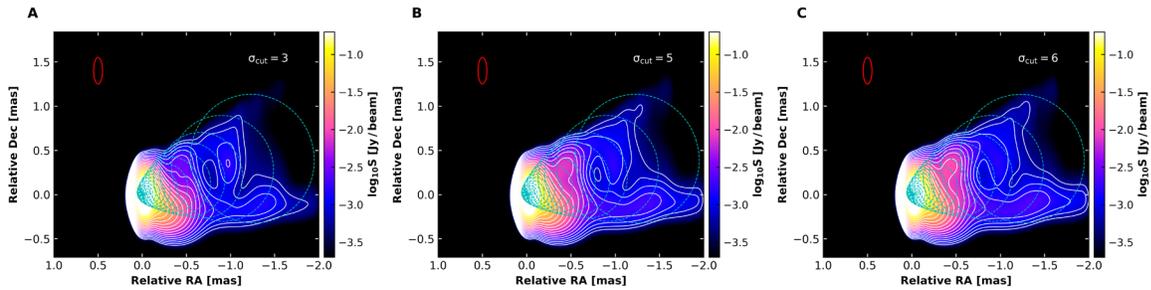

**Fig. S10. The predicted images at 86 GHz by MAD98 using different values of magnetization model parameter**. (**A**) $\sigma_{cut} = 3$. (**B**) $\sigma_{cut} = 5$. (**C**) $\sigma_{cut} = 6$. The three results are very similar, indicating that the modeling result is not sensitive to the choice of $\sigma_{cut}$. The dotted circles denote the boundary of the BZ-jet.

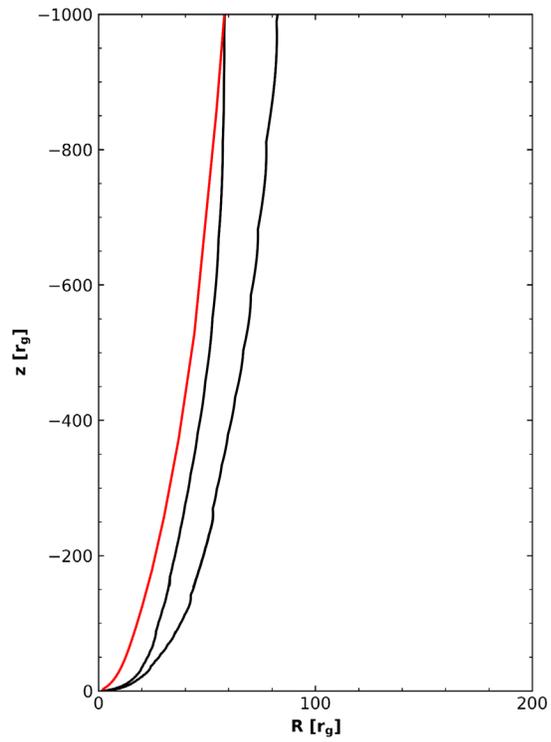

**Fig. S11. The comparison between the $\sigma_{cut} = cont.$ line and streamline based on $\phi$-averaged simulation data of MAD98.** The red line denotes the streamline, while the two black lines denote the $\sigma_{cut} = 5$ (closer to the jet axis) and $\sigma_{cut} = 1$ (away from the jet axis) lines.

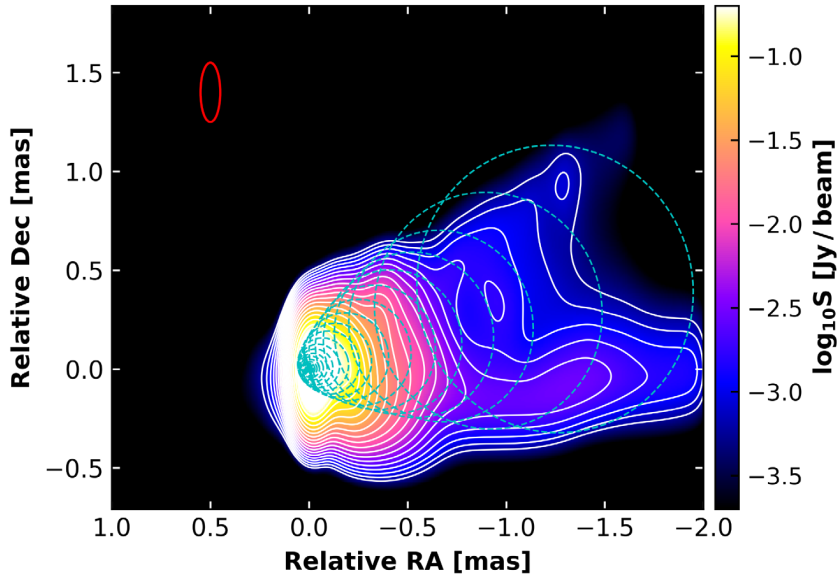

**Fig. S12. The predicted image at 86 GHz using the streamline as the boundary of the radiation region in the jet.** The streamline is the one shown in Fig. S11. This image is very similar to the image produced by the fiducial model shown by the top-middle panel of Fig. 1. The dotted circles denote the boundary of the BZ-jet.

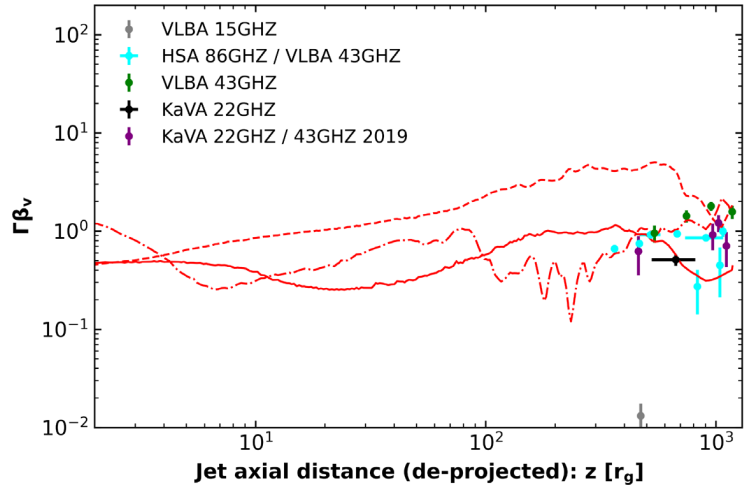

**Fig. S13. Comparison of the predicted (lines) and observed (dots) jet velocity as a function of de-projected distance from the core.** The dotted, dashed, and solid lines represent the distribution of $\Gamma\beta$ along the line that is located at 10%, 50%, and 100% of the outer boundary of the BZ-jet, respectively. The observational data are taken from Refs. (*67 & 68*) (VLBA 15 GHz), (*69*) (HSA 86GHZ/VLBA 43GHZ), (*70*) (KaVA 22GHz), (*71*) (VLBA 43GHz), and (72) (KaVA 22GHz/43GHz).

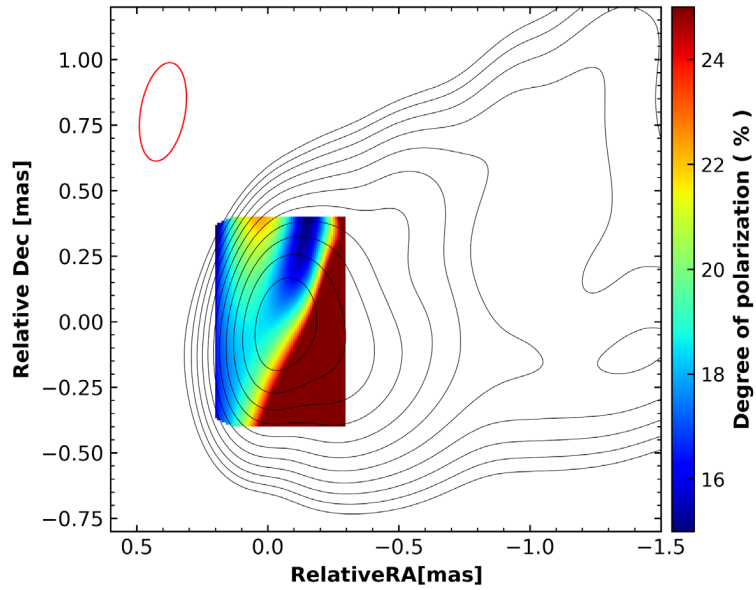

**Fig. S14. The polarization degree in the nuclear region of M87 predicted by the fiducial model.** For comparison with the observational result shown in Ref. (*73*), only the result in a central region is shown. Note that the predicted polarization degree shown here should be regarded as an upper limit when compared to the observational result.

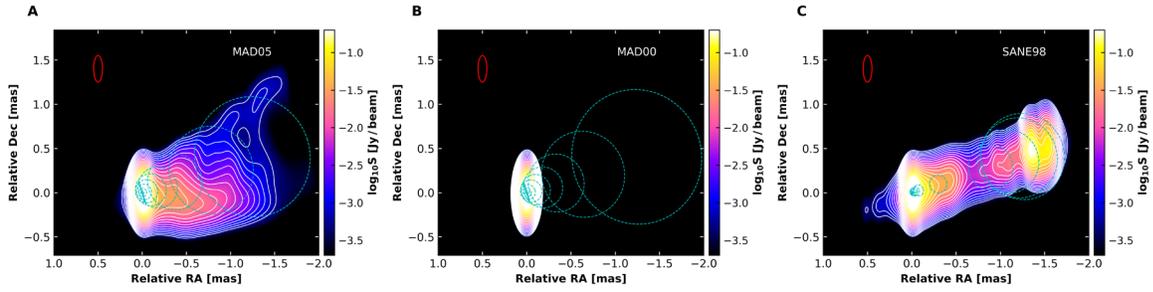

**Fig. S15. The jet images predicted by different models.** (**A**) MAD05, (**B**) MAD00, and (**C**) SANE98. The dotted circles denote the boundary of the BZ-jet.

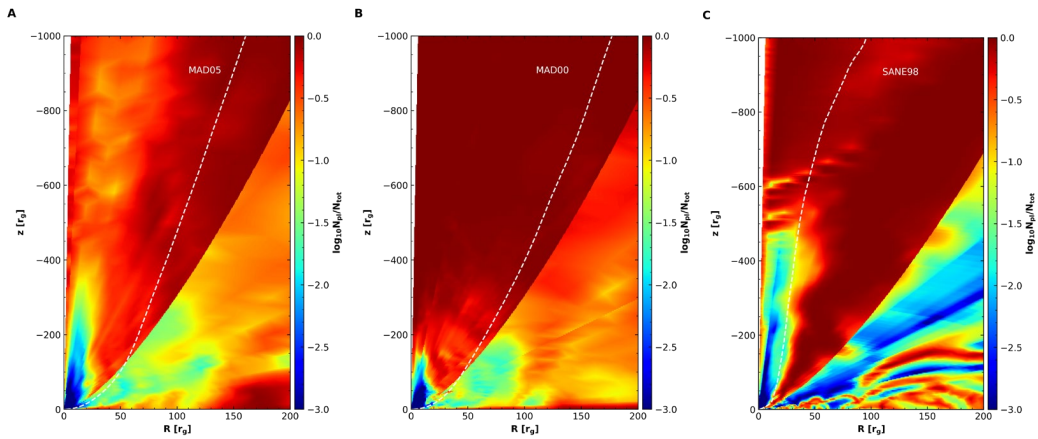

**Fig. S16. The $\phi$-averaged distribution of the ratio of nonthermal and total electrons number density predicted by different models.** They are (**A**) MAD05, (**B**) MAD00, and (**C**) SANE98, corresponding to Fig. S15. The white dashed line in all panels denotes the $\phi$-averaged boundary of the BZ-jet.

**Table S1.**

**The static mesh refinement grid of MAD98**

| Level | $r / r_g$ | $\theta / \pi$ | $\varphi / \pi$ |
|---|---|---|---|
| 0 | [1.1, 1200] | [0, 1] | [0, 2] |
| 1 | [1.1, 200] | [0.1305, 0.8689] | [0, 2] |
| 2 | [1.1, 30] | [0, 1] | [0, 2] |
| | [30, 1200] | [0.0722, 0.1667] | [0, 2] |
| | [30, 1200] | [0.8333, 0.9278] | [0, 2] |
| 3 | [30, 1200] | [0.0555, 0.0722] | [0, 2] |
| | [30, 1200] | [0.9278, 0.9444] | [0, 2] |
| | [30, 1200] | [0.0055, 0.0555] | [0, 2] |
| 4 | [30, 1200] | [0.9444, 0.9944] | [0, 2] |

**Table S2.**

**The static mesh refinement grid of SANE98, MAD05, and MAD00**

| Level | $r / r_g$ | $\theta / \pi$ | $\varphi / \pi$ |
|---|---|---|---|
| 0 | [1, 1200] | [0, 1] | [0, 2] |
| 1 | [1, 30] | [0, 0.1942] | [0, 2] |
|  | [1, 30] | [0.8058, 1] | [0, 2] |
|  | [1, 200] | [0.1305, 0.8695] | [0, 2] |
| 2 | [1, 30] | [0.1942, 0.8058] | [0, 2] |
|  | [6.4, 1200] | [0, 0.1209] | [0, 2] |
|  | [6.4, 1200] | [0.8791, 1] | [0, 2] |
| 3 | [36.7, 1200] | [0, 0.0573] | [0, 2] |
|  | [36.7, 1200] | [0.9427, 1] | [0, 2] |

**Table S3.**

**The mass accretion rates of various models**

| Model | $\eta$ | $\dot{M}$ |
|---|---|---|
| | | $(10^{-3}\,M_\odot/yr)$ |
| ``Thermal-only'' MAD98 | — | *0.103* |
| ``Current density'' MAD98 | *0.2 × 10⁻⁵* | 0.094 |
| ``Current density'' MAD05 | *1.3 × 10⁻⁵* | 0.398 |
| ``Current density'' MAD00 | *0.47 × 10⁻⁵* | 0.125 |
| ``Current density'' SANE98 | *0.008 × 10⁻⁵* | 0.374 |

**REFERENCE AND NOTES**

1. R. D. Blandford, R. L. Znajek, Electromagnetic extraction of energy from Kerr black holes. *Mon. Not. R. Astron. Soc.* **179,** 433–456 (1977).

2. J. F. Hawley, S. A. Balbus, The dynamical structure of nonradiative black hole accretion flows. *Astrophys. J.* **573**, 738–748 (2002).

3. A. Tchekhovskoy, R. Narayan, J. C. McKinney, Efficient generation of jets from magnetically arrested accretion on a rapidly spinning black hole. *Mon. Not. R. Astron. Soc.,* **418**, L79-L83 (2011).

4. J. C. McKinney, A. Tchekhovskoy, R. D. Blandford, General relativistic magnetohydrodynamic simulations of magnetically choked accretion flows around black holes. *Mon. Not. R. Astron. Soc.,* **423**, 3083–3117 (2012)

5. A. Sadowski, R. Narayan, R. Penna, Y. Zhu, Energy, momentum and mass outflows and feedback from thick accretion discs around rotating black holes. *Mon. Not. R. Astron. Soc.,* **436**, 3856–3874 (2013).

6. R. D. Blandford, D. G. Payne, Hydromagnetic flows from accretion discs and the production of radio jets. *Mon. Not. R. Astron. Soc.* **199,** 883–903 (1982)

7. D. Lynden-Bell, On why discs generate magnetic towers and collimate jets. *Mon. Not. R. Astron. Soc.,* **341**, 1360–1372 (2003).

8. Y. Kato, S. Mineshige, K. Shibata, Magnetohydrodynamic accretion flows: Formation of magnetic tower jet and subsequent quasi-steady state. *Astrophys. J.* **605**, 307–320 (2004).

9. J. F. Hawley, J. H. Krolik, Magnetically driven jets in the kerr metric. *Astrophys. J.* **641**, 103 (2006), 116.

10. F. Yuan, Z. Gan, R. Narayan, A. Sadowski, D. Bu, X-N. Bai, Numerical Simulation of Hot Accretion Flows. III. Revisiting Wind Properties Using the Trajectory Approach. *Astrophys. J.* **804,** 101 (2015).

11. Materials and methods are available at the end of the main text while supplementary materials are available at the Science Advances website.

12. Event Horizon Telescope Collaboration, K. Akiyama, J. C. Algaba, A. Alberdi, W. Alef, R. Anantua, K. Asada, R. Azulay, A-K. Baczko, D. Ball, M. Baloković, J. Barrett, B. A. Benson, D. Bintley, L. Blackburn, R. Blundell, W. Boland, K. L. Bouman, G. C. Bower, H. Boyce, M. Bremer, C. D. Brinkerink, R. Brissenden, S. Britzen, A. E. Broderick, D. Broguiere, T. Bronzwaer, D-Y. Byun, J. E. Carlstrom, A. Chael, C-K. Chan, S. Chatterjee, K. Chatterjee, M-T. Chen, Y. Chen, P. M. Chesler, I. Cho, P. Christian, J. E. Conway, J. M. Cordes, T. M. Crawford, G. B. Crew, A. Cruz-Osorio, Y. Cui, J. Davelaar, M. De Laurentis, R. Deane, J. Dempsey, G. Desvignes, J. Dexter, S. S. Doeleman, R. P. Eatough, H. Falcke, J. Farah, V. L. Fish, E. Fomalont, H. A. Ford, R. Fraga-Encinas, P. Friberg, C. M. Fromm, A. Fuentes, P. Galison, C. F. Gammie, R. García, Z. Gelles, O. Gentaz, B. Georgiev, C. Goddi, R. Gold, J. L. Gómez, A. I. Gómez-Ruiz, M. Gu, M. Gurwell, K. Hada, D. Haggard, M. H. Hecht, R. Hesper, E. Himwich, L. C. Ho, P. Ho, M. Honma, C-W. L. Huang, L. Huang, D. H. Hughes, S. Ikeda, M. Inoue, S. Issaoun, D. J. James, B. T. Jannuzi, M. Janssen, B. Jeter, W. Jiang, A. Jimenez-Rosales, M. D. Johnson, S. Jorstad, T. Jung, M. Karami, R. Karuppusamy, T. Kawashima, G. K. Keating, M. Kettenis, D-J. Kim, J-Y. Kim, J. Kim, J. Kim, M. Kino, J. Y. Koay, Y. Kofuji, P. M. Koch, S. Koyama, M. Kramer, C. Kramer, T. P. Krichbaum, C-Y. Kuo, T. R. Lauer, S-S. Lee, A. Levis, Y-R. Li, Z. Li, M. Lindqvist, R. Lico, G. Lindahl, J. Liu, K. Liu, E. Liuzzo, W-P. Lo, A. P. Lobanov, L. Loinard, C. Lonsdale, R-S. Lu, N. R. MacDonald, J. Mao, N. Marchili, S. Markoff, D. P. Marrone, A. P. Marscher, I. Martí-Vidal, S. Matsushita, L. D. Matthews, L. Medeiros, K. M. Menten, I. Mizuno, Y. Mizuno, J. M. Moran, K. Moriyama, M. Moscibrodzka, C. Müller, G. Musoke, M. Mus, Alejandro, D. Michalik, A. Nadolski, H. Nagai, N. M. Nagar, M. Nakamura, R. Narayan, G. Narayanan, I. Natarajan, A. Nathanail, J. Neilsen, R. Neri, C. Ni, A. Noutsos, M. Nowak, H. Okino, H. Olivares, G. N. Ortiz-León, T. Oyama, F. Özel, D. C. M. Palumbo, J. Park, N. Patel, U-L. Pen, D. W. Pesce, V. Piétu, R. Plambeck, A. PopStefanija, O. Porth, F. M. Pötzl, B. Prather, J. A. Preciado-López, D. Psaltis, H.-Y. Pu, V. Ramakrishnan, R. Rao, M. G. Rawlings, A. W. Raymond, L. Rezzolla, A. Ricarte, B. Ripperda, F. Roelofs, A. Rogers, E. Ros, M. Rose, A. Roshanineshat, H. Rottmann, A. L. Roy, C. Ruszczyk, K. L. J. Rygl, S. Sánchez, D. Sánchez-Arguelles, M. Sasada, T. Savolainen, F. P. Schloerb, K-F. Schuster, L. Shao, Z. Shen, D. Small, B. W. Sohn, J. SooHoo, H. Sun, F. Tazaki, A. J. Tetarenko, P. Tiede, R. P. J. Tilanus, M. Titus, K. Toma, P. Torne, T. Trent, E. Traianou, S. Trippe, I. van Bemmel, H. J. van Langevelde, D. R. van Rossum, J. Wagner, D. Ward-Thompson, J. Wardle, J. Weintroub, N. Wex, R.

Wharton, M. Wielgus, G. N. Wong, Q. Wu, D. Yoon, A. Young, K. Young, Z. Younsi, F. Yuan, Y-F. Yuan, J. A. Zensus, G-Y. Zhao, S-S. Zhao, First M87 Event Horizon Telescope Results. VIII. Magnetic Field Structure near The Event Horizon. *Astrophys. J.* **910,** L13 (2021).

13. F. Yuan, H. Wang, H. Yang, The accretion flow in M87 is really MAD. *Astrophys. J.* **924**, 124 (2022).

14. M. Mościbrodzka, H. Falcke, H. Shiokawa, General relativistic magnetohydrodynamical simulations of the jet in M 87. *Astron. Astrophys.* **586**, A38 (2016).

15. I.V. Igumenshchev, Magnetically arrested disks and the origin of poynting jets: A numerical study. *Astrphys. J.* **677**, 317–326 (2008).

16. J. Dexter, A. Tchekhovskoy, A. Jiménez-Rosales, S.M. Ressler, M. Baubӧck, Y. Dallilar, P. T. de Zeeuw, F. Eisenhauer, S. von Fellenberg, F. Gao, R. Genzel, S. Gillessen, M. Habibi, T. Ott, J. Stadler, O. Straub, F. Widmann, Sgr A* near-infrared flares from reconnection events in a magnetically arrested disc. *Mon. Not. R. Astron. Soc.* **497**, 4999–5007 (2020)

17. O. Porth, Y. Mizuno, Z. Younsi, C.M. Fromm, Flares in the Galactic Centre - I. Orbiting flux tubes in magnetically arrested black hole accretion discs. *Mon. Not. R. Astron. Soc.* **502**, 2023–2032 (2021).

18. B. Ripperda, M. Liska, K. Chatterjee, G. Musoke, A. A. Philippov, S. B. Markoff, A. Tchekhovskoy, Z. Younsi, Black hole flares: Ejection of accreted magnetic flux through 3D plasmoid-mediated reconnection. *Astrophys. J.* **924**, L32 (2022).

19. K. Chatterjee, R. Narayan, Flux eruption events drive angular momentum transport in magnetically arrested accretion flows. *Astrophys. J.* **941**, 30 (2022).

20. F. Guo, H. Li, W. Daughton, Y.-H. Liu, Formation of hard power laws in the energetic particle spectra resulting from relativistic magnetic reconnection. *Phys. Rev. Lett.*, **113**, 155005 (2014).

21. L. Sironi, A. Spitkovsky, Relativistic reconnection: An efficient source of non-thermal particles. *Astrophys. J.* **783**, L21 (2014).

22. D. Ball, L. Sironi, F. Özel, Electron and proton acceleration in trans-relativistic magnetic reconnection: Dependence on plasma beta and magnetization. *Astrophys. J.* **862**, 80 (2018)

23. G. R. Werner, D. A. Uzdensky, M. C. Begelman, B. Cerutti, K. Nalewajko, Non-thermal particle acceleration in collisionless relativistic electron-proton reconnection. *Mon. Not. R. Astron. Soc.* **473,** 4840–4861 (2018).

24. X. Li, F. Guo, Y.-H. Liu, H. Li, A model for nonthermal particle acceleration in relativistic magnetic reconnection. *Astrophys. J.* **954**, L37 (2023).

25. E. Petersen, C. Gammie, Non-thermal models for infrared flares from Sgr A*. *Mon. Not. R. Astron. Soc.* **494,** 5923–5935 (2020).

26. A. Cruz-Osorio, C. Fromm, Y. Mizuno, A. Nathanail, Z. Younsi, O. Porth, J. Davelaar, H. Falcke, M. Kramer, L. Rezzolla, State-of-the-art energetic and morphological modelling of the launching site of the M87 jet. *Nat. Astron.* **6**, 103–108 (2022).

27. M. Mościbrodzka, C. F. Gammie, IPOLE - semi-analytic scheme for relativistic polarized radiative transport. *Mon. Not. R. Astron. Soc.,* **475**, 43–54 (2018).

28. A. Chael, R. Narayan, M. D. Johnson, Two-temperature, magnetically arrested disc simulations of the jet from the supermassive black hole in M87. *Mon. Not. R. Astron. Soc.,* **486**, 2873–2895 (2019).

29. C. J. White, J. M. Stone, C. F. Gammie, An Extension of the Athena++ Code Framework for GRMHD Based on Advanced Riemann Solvers and Staggered-mesh Constrained Transport. *Astrophys. J., Suppl. Ser.* **225,** 22 (2016).

30. J. M. Stone, K. Tomida, C. J. White, K. G. Felker, The Athena++ Adaptive Mesh Refinement Framework: Design and Magnetohydrodynamic Solvers. *Astrophys. J., Suppl. Ser.* **249,** 4 (2020).

31. B. Einfeldt, On Godunov-Type Methods for Gas Dynamics. *SIAM J. Numer. Anal.* **25,** 294–318 (1988).

32. O. Porth, K. Chatterjee, R. Narayan, C. F. Gammie, Y. Mizuno, P. Anninos, J. G. Baker, M. Bugli, C. Chan, J. Davelaar, L. Del Zanna, Z. B. Etienne, P. C. Fragile, B. J. Kelly, M. Liska, S. Markoff, J. C.

McKinney, B. Mishra, S. C. Noble, H. Olivares, B. Prather, L. Rezzolla, B. R. Ryan, J. M. Stone, N. Tomei, C. J. White, Z. Younsi, K. Akiyama, A. Alberdi, W. Alef, K. Asada, R. Azulay, A.-K. Baczko, D. Ball, M. Baloković, J. Barrett, D. Bintley, L. Blackburn, W. Boland, K. L. Bouman, G. C. Bower, M. Bremer, C. D. Brinkerink, R. Brissenden, S. Britzen, A. E. Broderick, D. Broguiere, T. Bronzwaer, D.-Y. Byun, J. E. Carlstrom, A. Chael, S. Chatterjee, M.-T. Chen, Y. Chen, I. Cho, P. Christian, J. E. Conway, J. M. Cordes, Geoffrey, B. Crew, Y. Cui, M. De Laurentis, R. Deane, J. Dempsey, G. Desvignes, S. S. Doeleman, R. P. Eatough, H. Falcke, V. L. Fish, E. Fomalont, R. Fraga-Encinas, B. Freeman, P. Friberg, C. M. Fromm, J. L. Gómez, P. Galison, R. García, O. Gentaz, B. Georgiev, C. Goddi, R. Gold, M. Gu, M. Gurwell, K. Hada, M. H. Hecht, R. Hesper, L. C. Ho, P. Ho, M. Honma, C.-W. Huang, L. Huang, D. H. Hughes, S. Ikeda, M. Inoue, S. Issaoun, D. J. James, B. T. Jannuzi, M. Janssen, B. Jeter, W. Jiang, M. D. Johnson, S. Jorstad, T. Jung, M. Karami, R. Karuppusamy, T. Kawashima, G. K. Keating, M. Kettenis, J.-Y. Kim, J. Kim, J. Kim, M. Kino, J. Y. Koay, Patrick, M. Koch, S. Koyama, M. Kramer, C. Kramer, T. P. Krichbaum, C.-Y. Kuo, T. R. Lauer, S.-S. Lee, Y.-R. Li, Z. Li, M. Lindqvist, K. Liu, E. Liuzzo, W.-P. Lo, A. P. Lobanov, L. Loinard, C. Lonsdale, R.-S. Lu, N. R. MacDonald, J. Mao, D. P. Marrone, A. P. Marscher, I. Martí-Vidal, S. Matsushita, L. D. Matthews, L. Medeiros, K. M. Menten, I. Mizuno, J. M. Moran, K. Moriyama, M. Moscibrodzka, C. Müller, H. Nagai, N. M. Nagar, M. Nakamura, G. Narayanan, I. Natarajan, R. Neri, C. Ni, A. Noutsos, H. Okino, T. Oyama, F. Özel, D. C. M. Palumbo, N. Patel, U.-L. Pen, D. W. Pesce, V. Piétu, R. Plambeck, A. PopStefanija, J. A. Preciado-López, D. Psaltis, H.-Y. Pu, V. Ramakrishnan, R. Rao, M. G. Rawlings, A. W. Raymond, B. Ripperda, F. Roelofs, A. Rogers, E. Ros, M. Rose, A. Roshanineshat, H. Rottmann, A. L. Roy, C. Ruszczyk, K. L. J. Rygl, S. Sánchez, D. Sánchez-Arguelles, M. Sasada, T. Savolainen, F. P. Schloerb, K.-F. Schuster, L. Shao, Z. Shen, D. Small, B. W. Sohn, J. SooHoo, F. Tazaki, P. Tiede, R. P. J. Tilanus, M. Titus, K. Toma, P. Torne, T. Trent, S. Trippe, S. Tsuda, I. van Bemmel, H. J. van Langevelde, D. R. van Rossum, J. Wagner, J. Wardle, J. Weintroub, N. Wex, R. Wharton, M. Wielgus, G. N. Wong, Q. Wu, K. Young, A. Young, F. Yuan, Y.-F. Yuan, J. A. Zensus, G. Zhao, S.-S. Zhao, Z. Zhu, Event Horizon Telescope Collaboration, The event horizon general relativistic magnetohydrodynamic code comparison project. *Astrophys. J. Supp.* **243,** 26 (2019).

33. J. C. McKinney, C. F. Gammie, A Measurement of the electromagnetic luminosity of a Kerr black Hole. *Astrophys. J.* **611,** 977–995 (2004).

34. R. Narayan, I. V. Igumenshchev, M. A. Abramowicz, Magnetically arrested disk: An energetically efficient accretion flow. *Publ. Astron. Soc. Jpn.* **55,** L69–L72 (2003).

35. R. Narayan, A. SÄdowski, R. F. Penna, A. K. Kulkarni, GRMHD simulations of magnetized advection-dominated accretion on a non-spinning black hole: Role of outflows. *Mon. Not. R. Astron. Soc.* **426,** 3241–3259 (2012)

36. L. G. Fishbone, V. Moncrief, Relativistic fluid disks in orbit around Kerr black holes. *Astrophys. J.* **207,** 962–976 (1976).

37. R. F. Penna, A. Kulkarni, R. Narayan, A new equilibrium torus solution and GRMHD initial conditions. *Astron. Astrophys.* **559,** A116 (2013).

38. C. J. White, F. Chrystal, The effects of resolution on black hole accretion simulations of jets. *Mon. Not. R. Astron. Soc.* **498,** 2428–2439 (2020).

39. J. F. Hawley, X. Guan, J. H. Krolik, Assessing quantitative results in accretion simulations: From local to global. *Astrophys. J.* **738**, 84 (2011).

40. R.-S. Lu, K. Asada, T. P. Krichbaum, J. Park, F. Tazaki, H.-Y. Pu, M. Nakamura, A. Lobanov, K. Hada, K. Akiyama, J-Y. Kim, I. Marti-Vidal, J. L. Gomez, T. Kawashima, F. Yuan, E. Rose, W. Alef, S. Britzen, M. Bremer, P. T. P. Ho, M. Honma, D. H. Hughes, M. Inoue, W. Jiang, M. Kino, S. Koyama, M. Lindqvist, J. Liu, A. P. Marscher, S. Matsushita, H. Nagai, H. Rottmann, T. Savolainen, K-F. Schuster, Z-Q. Shen, P. de Vicente, R. C. Walker, H. Yang, J. A. Zensus, J. C. Algaba, A. Allardi, U. Bach, R. Berthold, D. Bintley, D-Y. Byun, C. Casadio, S-H. Chang, C-C. Chang, S-C. Chang, C-C. Chen, M-T. Chen, R. Chilson, T. C. Chuter, J. Conway, G. B. Crew, J. T. Dempsey, S. Dornbusch, A. Faber, P. Friberg, J. G. Garcia, M. G. Garrido, C-C. Han, K.-C. Han, Y. Hasegawa, R. Herrero-Illana, Y.-D. Huang, C.-W. L. Huang, V. Impellizzeri, H. Jiang, H. Jinchi, T. Jung, J. Kallunki, P. Kirves, K. Kimura, J. Y. Koay, P. M. Koch, C. Kramer, A. Kraus, D. Kubo, C.-Y. Kuo, C.-T. Li, L. C.-C. Lin, C.-T. Liu, K.-Y. Liu, W.-P. Lo, L.-M. Lu, N. MacDonald, P. Martin-Cocher, H. Messias, Z. Meyer-Zhao, A. Minter, D. G. Nair, H. Nishioka, T. J. Norton, G. Nystrom, H. Ogawa, P. Oshiro, N. A. Patel, U.-L. Pen, Y. Pidopryhora, N. Pradel, P. A. Raffin, R. Rao, I. Ruiz, S. Sanchez, P. Shaw, W. Snow, T. K. Sridharan, R. Srinivasan, B. Tercero, P. Torne, E. Traianou, J. Wagner, C. Walther, T-S. Wei, J. Yang,

C.-Y. Yu, A ring-like accretion structure in M87 connecting its black hole and jet. *Nature.* **616**, 686–690 (2023).

41. Q. Zhang, F. Guo, W. Daughton, H. Li, X. Li, Efficient nonthermal ion and electron acceleration enabled by the flux-rope kink instability in 3D nonrelativistic magnetic reconnection. *Phys. Rev. Lett.*, **127**, 185101 (2021).

42. F. Özel, D. Psaltis, R. Narayan, Hybrid thermal-nonthermal synchrotron emission from hot accretion flows. *Astrophys. J.* **541,** 234 (2000), 249.

43. F. Yuan, E. Quataert, R. Narayan, Nonthermal electrons in radiatively inefficient accretion flow models of Sagittarius A*. *Astrophys. J.* **598,** 301 (2003), 312.

44. J. Dexter, J. C. McKinney, E. Agol, The size of the jet launching region in M87. *Mon. Not. R. Astron. Soc.* **421,** 1517 (2012), 1528.

45. K. Chatterjee, S. Markoff, J. Neilsen, Z. Younsi, G. Witzel, A. Tchekhovskoy, D. Yoon, A. Ingram, M. van der Klis, H. Boyce, T. Do, D. Haggard, M. A. Nowak, General relativistic MHD simulations of non-thermal flaring in Sagittarius A*. *Mon. Not. R. Astron. Soc.,* **507**, 5281–5302 (2021).

46. A. E. Broderick, V. L. Fish, M. D. Johnson, K. Rosenfeld, C. Wang, S. S. Doeleman, K. Akiyama, T. Johannsen, A. L. Roy, Modeling seven years of event horizon telescope observations with radiatively inefficient accretion flow models. *Astrophys. J.* **820,** 137 (2016).

47. J. Davelaar, M. Mościbrodzka, T. Bronzwaer, H. Falcke, General relativistic magnetohydrodynamicalκ-jet models for Sagittarius A*. *Astron. Astrophys.* **612,** A34 (2018).

48. A. E. Broderick, J. C. McKinney, Parsec-scale faraday rotation measures from general relativistic magnetohydrodynamic simulations of active galactic nucleus jets. *Astrophys. J.* **725,** 773 (2010).

49. J. Davelaar, H. Olivares, O. Porth, T. Bronzwaer, M. Janssen, F. Roelofs, Y. Mizuno, C. M. Fromm, H. Falcke, L. Rezzolla, Modeling non-thermal emission from the jet-launching region of M 87 with adaptive mesh refinement. *Astron. Astrophys.* **632,** A2 (2019).

50. D. Ball, F. Özel, D. Psaltis, C. K. Chan, L. Sironi, The properties of reconnection current sheets in GRMHD simulations of radiatively inefficient accretion flows, *Astrophys. J.* **853**, 184 (2018).

51. C. B. Singh, Y. Mizuno, E. M. de Gouveia Dal Pino, Spatial growth of current-driven instability in relativistic rotating jets and the search for magnetic reconnection. *Astrophys. J.* **824**, 48 (2016).

52. E. P. Alves, J. Zrake, F. Fiuza, efficient nonthermal particle acceleration by the kink instability in relativistic jets. *Phys. Rev. Lett.*, **121**, 245101 (2018)

53. T. E. Medina-Torrejón, E. M. de Gouveria Dal Pino, L. H. S. Kadowaki, G. Kowal, C. B. Singh, Y. Mizuno, Particle acceleration by relativistic magnetic reconnection driven by kink instability turbulence in poynting flux-dominated jets. *Astrophys. J.* **908**, 193 (2021).

54. A. Tomimatsu, T. Matsuoka, M. Takahashi, Screw instability in black hole magnetospheres and a stabilizing effect of field-line rotation. *Phys.Rev. D.* **64**, 123003 (2001).

55. M. Nakamura, H. Li, S. T. Li, Stability properties of magnetic tower jets. *Astrophys. J.* **656,** 721–732 (2007).

56. H. Yang, F. Yuan, Y. F. Yuan, C. White, Numerical simulation of hot accretion flows (IV): Effects of black hole spin and magnetic field strength on the wind and the comparison between wind and jet properties. *Astrophys. J.* **914,** 131 (2021).

57. K. Nishikawa, Y. Mizuno, J. L. Gomez, I. Dutan, J. Niemiec, O. Kobzar, N. MacDonald, A. Meli, M. Pohl, K. Hirotani, Rapid particle acceleration due to recollimation shocks and turbulent magnetic fields in injected jets with helical magnetic fields. *Mon. Not. R. Astron. Soc.* **493,** 2652–2658 (2020).

58. L. Sironi, M. E. Rowan, R. Narayan, Reconnection-driven particle acceleration in relativistic shear flows. *Astrophys. J.* **907**, L44 (2021).

59. B. R. Ryan, S. M. Ressler, J. C. Dolence, C. Gammie, E. Quataert, Two-temperature GRRMHD simulations of M87. *Astrophys. J.* **864,** 126 (2018).

60. Event Horizon Telescope Collaboration, K. Akiyama, A. Alberdi, W. Alef, K. Asada, R. Azulay, A.-K. Baczko, D. Ball, M. Baloković, J. Barrett, D. Bintley, L. Blackburn, W. Boland, K. Bouman, G. C.

Bower, M. Bremer, C. D. Brinkerink, R. Brissenden, S. Britzen, A. E. Broderick, D. Broguiere, T. Bronzwaer, D.-Y. Byun, J. E. Carlstrom, A. Chael, C. Chan, S. Chatterjee, K. Chatterjee, M.-T. Chen, Y. Chen, I. Cho, P. Christian, J. E. Conway, J. M. Cordes, G. B. Crew, Y. Cui, J. Davelaar, M. De Laurentis, R. Deane, J. Dempsey, G. Desvignes, J. Dexter, S. S. Doeleman, R. P. Eatough, H. Falcke, V. L. Fish, E. Fomalont, R. Fraga-Encinas, P. Friberg, C. M. Fromm, J. L. Gómez, P. Galison, C. F. Gammie, R. García, O. Gentaz, B. Georgiev, C. Goddi, R. Gold, M. Gu, M. Gurwell, K. Hada, M. H. Hecht, R. Hesper, L. C. Ho, P. Ho, M. Honma, C.-W. L. Huang, L. Huang, D. H. Hughes, S. Ikeda, M. Inoue, S. Issaoun, D. J. James, B. T. Jannuzi, M. Janssen, B. Jeter, W. Jiang, M. D. Johnson, S. Jorstad, T. Jung, M. Karami, R. Karuppusamy, T. Kawashima, G. K. Keating, M. Kettenis, J.-Y. Kim, J. Kim, J. Kim, M. Kino, J. Y. Koay, P. M. Koch, S. Koyama, M. Kramer, C. Kramer, T. P. Krichbaum, C.-Y. Kuo, T. R. Lauer, S.-S. Lee, Y.-R. Li, Z. Li, M. Lindqvist, K. Liu, E. Liuzzo, W.-P. Lo, A. P. Lobanov, L. Loinard, C. Lonsdale, R.-S. Lu, N. R. MacDonald, J. Mao, S. Markoff, D. P. Marrone, A. P. Marscher, I. Martí-Vidal, S. Matsushita, L. D. Matthews, L. Medeiros, K. M. Menten, Y. Mizuno, I. Mizuno, J. M. Moran, K. Moriyama, M. Moscibrodzka, C. Muĺler, H. Nagai, N. M. Nagar, M. Nakamura, R. Narayan, G. Narayanan, I. Natarajan, R. Neri, C. Ni, A. Noutsos, H. Okino, H. Olivares, T. Oyama, F. Özel, D. C. M. Palumbo, N. Patel, U.-L. Pen, D. W. Pesce, V. Piétu, R. Plambeck, A. PopStefanija, O. Porth, B. Prather, J. A. Preciado-López, D. Psaltis, H.-Y. Pu, V. Ramakrishnan, R. Rao, M. G. Rawlings, A. W. Raymond, L. Rezzolla, B. Ripperda, F. Roelofs, A. Rogers, E. Ros, M. Rose, A. Roshanineshat, H. Rottmann, A. L. Roy, C. Ruszczyk, B. R. Ryan, K. L. J. Rygl, S. Sánchez, D. Sánchez-Arguelles, M. Sasada, T. Savolainen, F. P. Schloerb, K.-F. Schuster, L. Shao, Z. Shen, D. Small, B. W. Sohn, J. SooHoo, F. Tazaki, P. Tiede, R. P. J. Tilanus, M. Titus, K. Toma, P. Torne, T. Trent, S. Trippe, S. Tsuda, I. van Bemmel, H. J. van Langevelde, D. R. van Rossum, J. Wagner, J. Wardle, J. Weintroub, N. Wex, R. Wharton, M. Wielgus, G. N. Wong, Q. Wu, A. Young, K. Young, Z. Younsi, F. Yuan, Y.-F. Yuan, J. A. Zensus, G. Zhao, S.-S. Zhao, Z. Zhu, J. Anczarski, F. K. Baganoff, A. Eckart, J. R. Farah, D. Haggard, Z. Meyer-Zhao, D. Michalik, A. Nadolski, J. Neilsen, H. Nishioka, M. A. Nowak, N. Pradel, R. A. Primiani, K. Souccar, L. Vertatschitsch, P. Yamaguchi, S. Zhang, First M87 event horizon telescope results. V. physical origin of the asymmetric ring. *Astrophys. J.* **875**, L5 (2019).

61. K. Gebhardt, J. Adams, D. Richstone, T. R. Lauer, S. M. Faber, K. Gultekin, J. Murphy, S. Tremaine, The black hole mass in M87 from Gemini/NIFS adaptive optics observations. *Astrophys. J.* **729,** 119 (2011).

62. R. C. Walker, P. E. Hardee, F. B. Davies, C. Ly, W. Junor, The structure and dynamics of the subparsec jet in M87 based on 50 VLBA observations over 17 years at 43 GHz. *Astrophys. J.* **855,** 128 (2018).

63. K. Asada, M. Nakamura, The structure of the M87 jet: A transition from parabolic to conical streamlines. *Astrophys. J.* **745**, L28 (2012).

64. J.-Y. Kim, T. P. Krichbaum, R.-S. Lu, E. Ros, U. Bach, M. Bremer, P. de Vicente, M. Lindqvist, J. A. Zensus, The limb-brightened jet of M87 down to the 7 Schwarzschild radii scale. *Astron. Astrophys.* **616,** A188 (2018).

65. M. Janssen, C. Goddi, I. M. van Bemmel, M. Kettenis, D. Small, E. Liuzzo, K. Rygl, I. Marti-Vidal, L. Blackburn, M. Wielgus, H. Falcke, rPICARD: A CASA-based calibration pipeline for VLBI data. Calibration and imaging of 7 mm VLBA observations of the AGN jet in M 87. arXiv: 1902.01749 [astro-ph.IM] (15 May 2019).

66. M. Nakamura, K. Asada, K. Hada, H.-Y. Pu, S. Noble, C. Tseng, K. Toma, M. Kino, H. Nagai, K. Takahashi, J-C Algaba, M. Orienti, K. Akiyama, A. Doi, G. Giovannini, M. Giroletti, M. Honma, S. Koyama, R. Lico, K. Niinuma, F. Tazaki, Parabolic Jets from the Spinning Black Hole in M87. *Astrophys. J.* **868**, 146 (2018).

67. K. I. Kellermann, M. L. Lister, D. C. Homan, R. C. Vermeulen, M. H. Cohen, E. Ros, M. Kadler, J. A. Zensus, Y. Y. Kovalev, Sub-milliarcsecond imaging of quasars and active galactic nuclei. III. kinematics of parsec-scale radio jets. *Astrophys. J.* **609**, 539–563 (2004).

68. Y. Y. Kovalev, M. L. Lister, D. C. Homan, K. I. Kellermann, The inner jet of the radio galaxy M87. *Astrophys. J.* **668**, L27-L30 (2007).

69. K. Hada, M. Kino, A. Doi, H. Nagai, M. Honma, K. Akiyama, F. Tazaki, R. Lico, M. Giroletti, G. Giovannini, M. Orienti, Y. Hagiwara, High-sensitivity 86 GHz (3.5 mm) VLBI observations of M87: Deep imaging of the jet base at a resolution of 10 Schwarzschild radii. *Astrophys. J.* **817**, 131 (2016).

70. K. Hada, J. H. Park, M. Kino, K. Niinuma, B. W. Sohn, H. W. Ro, T. Jung, J.-C. Algaba, G.-Y. Zhao, S.-S. Lee, K. Akiyama, S. Trippe, K. Wajima, S. Sawada-Satoh, F. Tazaki, I. Cho, J. Hodgson, J. A. Lee, Y. Hagiwara, M. Honma, S. Koyama, J. Oh, T. Lee, H. Yoo, N. Kawaguchi, D-G. Roh, S.-J. Oh,

J.-H. Yeom, D.-K. Jung, C. Oh, H.-R. Kim, J.-Y. Hwang, D.-Y. Byun, S.-H. Cho, H.-G. Kim, H. Kobayashi, K. M. Shibata, Pilot KaVA monitoring on the M 87 jet: Confirming the inner jet structure and superluminal motions at sub-pc scales. *Publ. Astron. Soc. Jpn.*, **69**, 71 (2017).

71. F. Mertens, A. P. Lobanov, R. C. Walker, P. E. Hardee, Kinematics of the jet in M 87 on scales of 100-1000 Schwarzschild radii. *Astron. Astrophys.* **595**, A54 (2016).

72. J. Park, K. Hada, M. Kino, M. Nakamura, J. Hodgson, H. Ro, Y. Cui, K. Asada, J-C. Algaba, S. Sawada-Satoh, S-S. Lee, I. Cho, Z. Shen, W. Jiang, S. Trippe, K. Niinuma, B. W. Sohn, T. Jung, G.-Y. Zhao, K. Wajima, F. Tazaki, M. Honma, T. An, K. Akiyama, D.-Y. Byun, J. Kim, Y. Zhang, X. Cheng, H. Kobayashi, K. M. Shibata, J. W. Lee, D.-G. Roh, S.-J. Oh, J.-H. Yeom, D.-K. Jung, C. Oh, H.-R. Kim, J.-Y. Hwang, Y. Hagiwara, Kinematics of the M87 Jet in the collimation zone: Gradual acceleration and velocity stratification. *Astrophys. J.* **887**, 147 (2019).

73. E. Kravchenko, M. Giroletti, K. Hada, D. L. Meier, M. Nakamura, J. Park, R. C. Walker, Linear polarization in the nucleus of M87 at 7 mm and 1.3 cm. *Astron. Astrophys.* **637,** L6 (2020).

74. J. Park, K. Hada, M. Kino, M. Nakamura, H. Ro, and S. Trippe, Faraday rotation in the Jet of M87 inside the Bondi radius: Indication of winds from hot accretion flows confining the relativistic jet. *Astrophys. J.* **871**, 257 (2019).

75. V. S. Beskin, Ya. N. Istomin, and V. I. Parev, Filling the magnetosphere of a supermassive black hole with plasma. *Sov. Astron.* **36**, 642 (1992).

76. A. Levinson, F. Rieger, Variable TeV EMISSION as a manifestation of jet formation in M87?. *Astrophys. J.* **730**, 123 (2011).

77. A. E. Broderick, A. Tchekhovskoy, Horizon-scale lepton acceleration in jets: Explaining the compact radio emission in M87. *Astrophys. J.* **809**, 97 (2015).

78. A. A. Zdziarski, D. G. Phuravhathu, M. Sikora, M. Böttcher, J. O. Chibueze, The composition and power of the jet of the broad-line radio galaxy 3C 120. *Astrophys. J.* **928**, L9 (2022).

79. E. E. Nokhrina, V. S. Beskin, Y. Y. Kovalev, A. A. Zheltoukhov, Intrinsic physical conditions and structure of relativistic jets in active galactic nuclei. *Mon. Not. R. Astron. Soc.* **447**, 2726–2737 (2015).